\documentclass[%
 aip,
 amsmath,amssymb,
 reprint,%
]{revtex4-1}

\usepackage{graphicx}
\usepackage{dcolumn}
\usepackage{bm}
\usepackage{rotating}
\usepackage[section]{placeins}

\usepackage[utf8]{inputenc}
\usepackage[T1]{fontenc}
\usepackage{mathptmx}

\begin{document}

\preprint{AIP/123-QED}


\title[S. Mumford et al.]{Sample Shape and Boundary Dependence of Measured Transverse Thermal Properties}

\author{Samuel Mumford}
\affiliation{Geballe Laboratory for Advanced Materials, Stanford University, Stanford CA, 94305, USA}
\affiliation{Department of Physics, Stanford University, Stanford CA, 94305, USA}
\author{Tiffany Paul}%
\affiliation{Geballe Laboratory for Advanced Materials, Stanford University, Stanford CA, 94305, USA}
\affiliation{Department of Applied Physics, Stanford University, Stanford CA, 94305, USA}
\author{Erik Kountz}
\affiliation{Geballe Laboratory for Advanced Materials, Stanford University, Stanford CA, 94305, USA}
\affiliation{Department of Physics, Stanford University, Stanford CA, 94305, USA}
\author{Aharon Kapitulnik}%
\affiliation{Geballe Laboratory for Advanced Materials, Stanford University, Stanford CA, 94305, USA}
\affiliation{Department of Physics, Stanford University, Stanford CA, 94305, USA}
\affiliation{Department of Applied Physics, Stanford University, Stanford CA, 94305, USA}%

\date{\today}

\begin{abstract}
Despite increased interest in thermal Hall measurements for the analysis of insulating quantum materials, there remains large uncertainty in such measurements due to contact misalignment. In this paper we propose that sample geometry and uncertain boundary conditions may account for uncertainty in the measurement of $D_{xy}$ or $\kappa_{xy}$ as well. By running simple simulations in an open source finite-element solver, we demonstrate that measured $D_{xy}$ in a thermal Hall bar can be changed by a factor of order unity in samples with similar width and length. This geometric corrective factor depends on the distinction between uniform heat flow and constant temperature boundary couplings to a bath. Sample geometry and boundary conditions can be accounted for through simulation or by using rectangular samples to make thermal Hall measurements more reliable and reproducible when the amplitude of $\kappa_{xy}$ is important. Finally, we propose a contactless optical method for measuring $D_{xy}$ which is insensitive to the longitudinal diffusivity pollution caused by contact misalignment. \footnote{Corresponding author is Samuel Mumford at smumfor2@stanford.edu}
\end{abstract}
\maketitle

\section{Introduction}

Measurement of the thermal Hall effect has emerged as a unique tool to investigate the nature of elementary excitations in solids, especially in response to the application of a magnetic field. Unlike the electric Hall response, a longitudinal temperature gradient may couple to a host of elementary excitations in both conductors and insulators such as spin\cite{Ong} or non-quasiparticle response\cite{Yamashita}. Indeed, thermal Hall conductivity ($\kappa_{xy}$) measurements have become important to investigating magnetic insulators such as disordered or frustrated magnets and spin liquids\cite{Ong, Yamashita, Watanabe2016, kasahara_2018,kasahara_PRL, Doki_2018, Hentrich_2019} or magnons in kagome and pyrochlore lattices\cite{Hirschberger2015,Ideue2012,Onose2010}. Phonons also respond to a temperature gradient\cite{Taldenkov, Wyder, Yamashita} and may yield an unexpectedly large $\kappa_{xy}$\cite{Behnia2020}.

\begin{figure*}
\centering
\includegraphics[width=\textwidth]{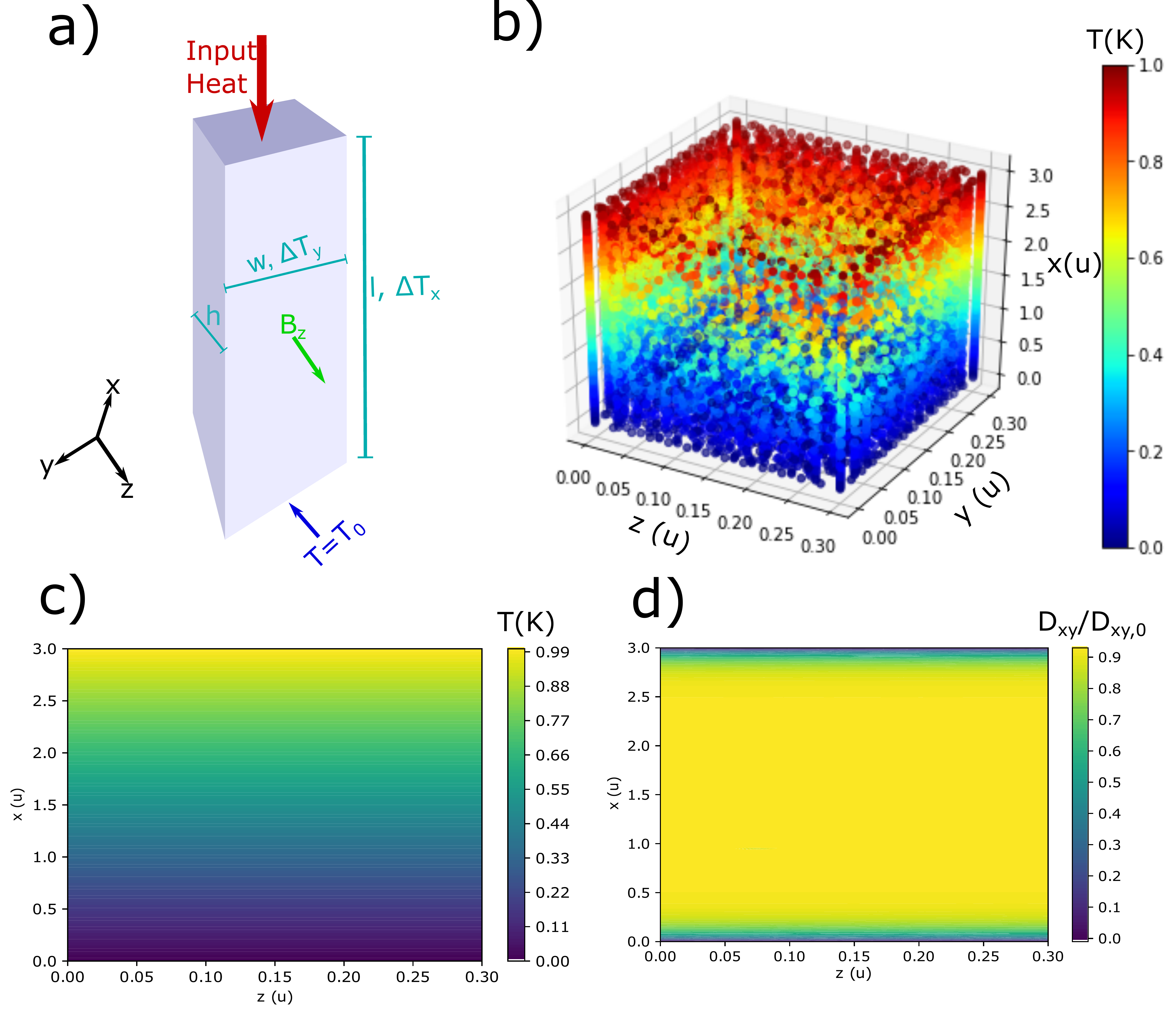}
\caption{ \label{setup} a) The geometry of the diffusivity simulation setup. Heat or constant nonzero temperature is applied to the top surface while the bottom surface is thermally anchored at $T=T_0$, generating a longitudinal thermal gradient in the $x$-direction. b) Temperature on each point in the FEniCS mesh due to applied longitudinal heat flow. c) Temperature projected onto the near $y$-surface. d) The ratio of the difference in temperature between opposing $y$-surfaces ($\Delta T_{y}$) and Eqn. \ref{baseEq} with set top and bottom surface temperatures. Note that the simulated and expected values do not agree near the temperature-controlled regions.}
\end{figure*}

Large uncertainties and discrepancies in transverse thermal transport measurement remain despite increased interest. For a thermal Hall bar designed as in Fig. \ref{setup}a, if the contacts used to measure transverse temperature difference $\Delta T_{y}$ are misaligned, the longitudinal thermal gradient $\Delta T_{x}$ pollutes measurement of $\kappa_{xy}$ or $D_{xy}$. This alignment uncertainty can account for an $\sim 20-50\%$ error in $\kappa_{xy}$ after antisymmetrization with magnetic field\cite{Behnia2020, Ong} and limits measurement of small $\kappa_{xy}$ much as it does for small $\rho_{xy}$. To correct for such misalignment error, we propose a new technique for measuring $D_{xy}$ optically based on an existing photothermal measurement technique for $\textbf{D}$\cite{Zhang5378}\cite{Zhang2019}.

Beyond this familiar source of uncertainty, the temperature anchoring of thermal Hall bars introduces a second source of measurement error not seen in voltage Hall bars. The differences in $\kappa_{xy}$ between ostensibly similar samples can be much larger than the statistical spread in data points, despite following similar temperature profiles\cite{Ong, Doki_2018}. 
In this paper, we propose through finite element simulation that such discrepancies may be attributed to geometric effects arising from the thermal boundary conditions. Such effects can change the measured $\kappa_{xy}$ by $\sim 50\%$ for square samples. As the geometric corrective factor depends only on one unitless parameter, thermal Hall measurements could be made with increased confidence by accounting for sample geometry either with simulation or experimental design changes.

\section{General Simulation Approach}

The heat diffusion equation with diffusivity $\textbf{D}$ can be written in frequency space for a heat source $g$ modulated at frequency $\omega$ as
\begin{equation}\label{baseTrue}
    i\omega T - \vec{\nabla}\cdot\left(\textbf{D}\vec{\nabla}T\right) = g(r).
\end{equation}
In order to address the effect of finite sample size on measurements of $\textbf{D}$, the temperature response of a sample was simulated using the open source finite element solver FEniCS\cite{fenics,dolfin}. FEniCS allows for the implementation of linear differential equations in the weak form. Using test functions $v_{1}$ and $v_{2}$, separating the real and imaginary parts of $T$ and $g$, and integrating by parts in Eqn. \ref{baseTrue} yields
\begin{equation}
    \int -\omega T_{i}v_{1} + \left(\textbf{D}\vec{\nabla}T_{r}\right)\cdot \vec{\nabla}v_{1} + \omega T_{r}v_{2} + \left(\textbf{D}\vec{\nabla}T_{i}\right)\cdot \vec{\nabla}v_{2} dV
\end{equation}
\begin{equation*}
    = \int g_{r}v_{1} + g_{i}v_{2} dV
\end{equation*}
and boundary terms. Note that for a DC measurement, $\omega = T_{i} = g_{i} = 0$, giving
\begin{equation}\label{SE}
    \int  \left(\textbf{D}\vec{\nabla}T_{r}\right)\cdot \vec{\nabla}v_{1} dV = \int g_{r}v_{1} dV.
\end{equation}

Boundary conditions can be enforced implicitly or hard-coded into FEniCS simulation. Constant temperature boundary conditions are enforced through setting $v_{1} = 0$ on the boundaries with set temperature. Heat flow $q(s)$ provided on the surface $S$ can be added to Eqn. \ref{SE} as
\begin{equation}\label{baseSim}
    \int  \left(\textbf{D}\vec{\nabla}T_{r}\right)\cdot \vec{\nabla}v_{1} dV = \int g_{r}v_{1} dV + \int q(s)v_{1} dS.
\end{equation}
Note that Eqn. \ref{baseSim} is equivalent to modeling electrical transport or a voltage Hall bar up to changing notation. The difference lies in shifting from the boundary conditions of a current source to those of anchoring to a heat bath. Such boundary effects in thermal Hall bars can be analyzed through solving the modified Poisson equation of Eqn. \ref{baseSim} with varying boundary conditions.

\section{Thermal Hall Bar Simulation}

For DC heat flow in the $x$-direction\cite{Taillefer2016}, 
\begin{equation}
    \frac{\frac{dT}{dy}}{\frac{dT}{dx}} = \frac{D_{xy}}{D_{xx}}.
\end{equation}
Assuming uniform and constant heat flow, for a sample desgned as in Fig. \ref{setup} with $y$-dimensional or transverse width $w$, $x$-dimensional or longitudinal length $l$, and $z$-dimensional or magnetic flux-direction height $h$, $dT/dy = \Delta T_{y}/w$ and $dT/dx = \Delta T_{x}/l$. The $h$-independent term is then
\begin{equation}\label{baseEq}
    \frac{\Delta T_{y}}{\Delta T_{x}} = \frac{w}{l}\frac{D_{xy}}{D_{xx}}.
\end{equation}

To assess the effect of boundary uncertainty in the application of Eqn. \ref{baseEq}, transverse thermal transport was simulated under the boundary conditions of:
\begin{itemize}
    \item Uniform heat flow on the top and bottom surfaces. This is the standard assumption in a thermal Hall bar leading to Eqn. \ref{baseEq}.
    \item Constant temperature boundary conditions for the top and bottom surfaces. This may be more accurate physically if a sample is directly attached to a temperature controlled substrate.
    \item Constant heat flow on the top surface and constant temperature on the bottom surface. This is analogous to a resistive heater on one surface and temperature anchoring to a cold bath on the opposing surface.
\end{itemize}

\subsection{Constant Heat Flow Boundary Conditions}
\begin{figure}
\centering
\includegraphics[width=.45\textwidth]{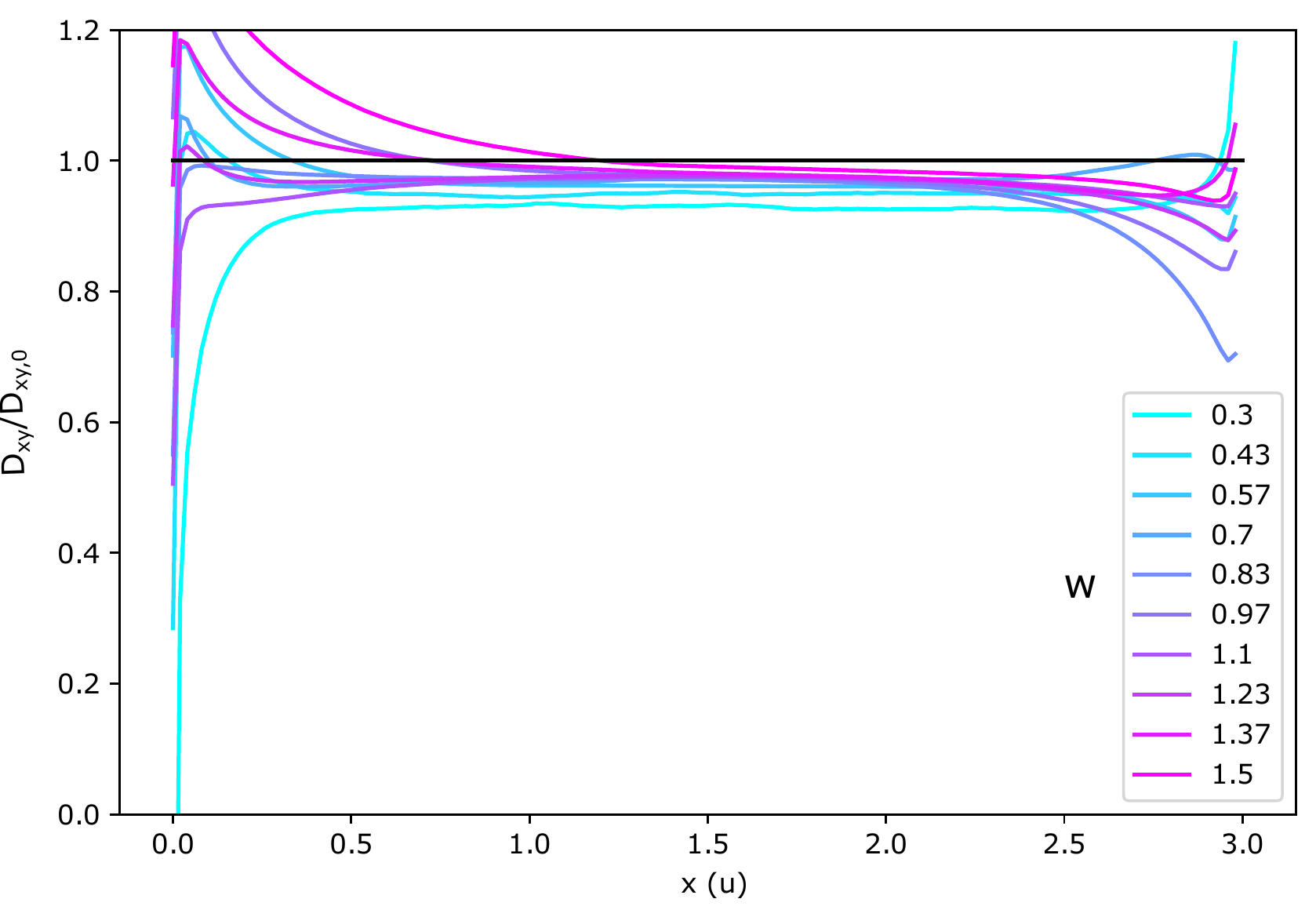}
\caption{ \label{BS}  Profile of the ratio of observed $D_{xy}$ to $D_{xy, 0} = 0.001~$u$^{2}$/s with uniform heat flow on the top and bottom surfaces at $x=0~$u and $x=3~$u depending on the position $x$ of $\Delta T_{y}$ measurement. Note that outside of the top and bottom $10\%$ of the sample, the measured and true values match to within the simulation uncertainty. Simulation is performed with $D_{xx} = D_{yy} = D_{zz} = 1~$u$^2$/s and $h = 0.2~$u.}
\end{figure}

As seen in Fig. \ref{BS}, the simulated $D_{xy}$ values match the true value outside of the extreme sample boundaries when the standard assumed boundary conditions of uniform heat flow are enforced. The edge values are strongly asymmetric on the top and bottom surfaces and depend on simulation mesh size. Such dependence on simulation parameters suggests that deviations from the expected behavior arise from simulation error and that agreement with Eqn. \ref{baseEq} would improve with increased computing power. 

Note that simulating Eqn. \ref{baseSim} is analogous to simulating transport in an electrical Hall bar in which the electrical analogue of Eqn. \ref{baseEq} is known to produce reliable results. Correspondingly, demonstrating that Eqn. \ref{baseEq} is reproduced in simulation with uniform heat flow is a necessary step in establishing simulation reliability. The observed agreement demonstrates that the FEniCS simulations reproduce the known experimental behavior of $D_{xy}$ in a Hall bar to within 10$\%$ outside of extrema and benchmarks simulation accuracy.

\subsection{Constant Temperature Boundary Conditions}

Experimental thermal Hall bars may not respect the assumed boundary conditions of uniform heat flow used for Eqn. \ref{baseEq}. Instead, the top surface is placed in contact with a heater and the bottom surface is anchored to a temperature-controlled substrate or heat sink\cite{Ong, Yamashita}. The method of such anchoring can vary from deposited patterned metal pads
\cite{Taillefer2016}, to grease covering the full bottom of the sample\cite{Behnia2020, Doki_2018}, to small wires\cite{kim_2019,Hirschberger2015}, to mechanical contact\cite{Wyder, Taldenkov}, and often is not specified\cite{kasahara_2018, kasahara_PRL, Watanabe2016, Yamashita, Hentrich_2019,Ideue2012,Onose2010}. Such thermal Hall bars therefore may exhibit constant temperature boundary conditions or highly non-uniform heat flow near the boundary. Such boundary conditions are incompatible with Eqn. \ref{baseEq}. As an example, if the bottom surface is anchored at $T = T_{0}$, there can be no $y-$dependence in the temperature and $\Delta T_{y} = 0$. Similarly, if the top surface is anchored to $T = T_{0} + \Delta T_{x}$, $\Delta T_{y} = 0$ on the top. As Eqn. \ref{baseEq} cannot apply at the sample boundaries at $x=0$ and $x=l$, simulations can be used to determine the viable $x$-range for $D_{xy}$ measurement with constant temperature boundary conditions. Outside of that range, there is a large thermometer placement-based error which would cause observed $D_{xy}$ to differ from the true value $D_{xy, 0}$.

\begin{figure*}
\centering
\includegraphics[width=\textwidth]{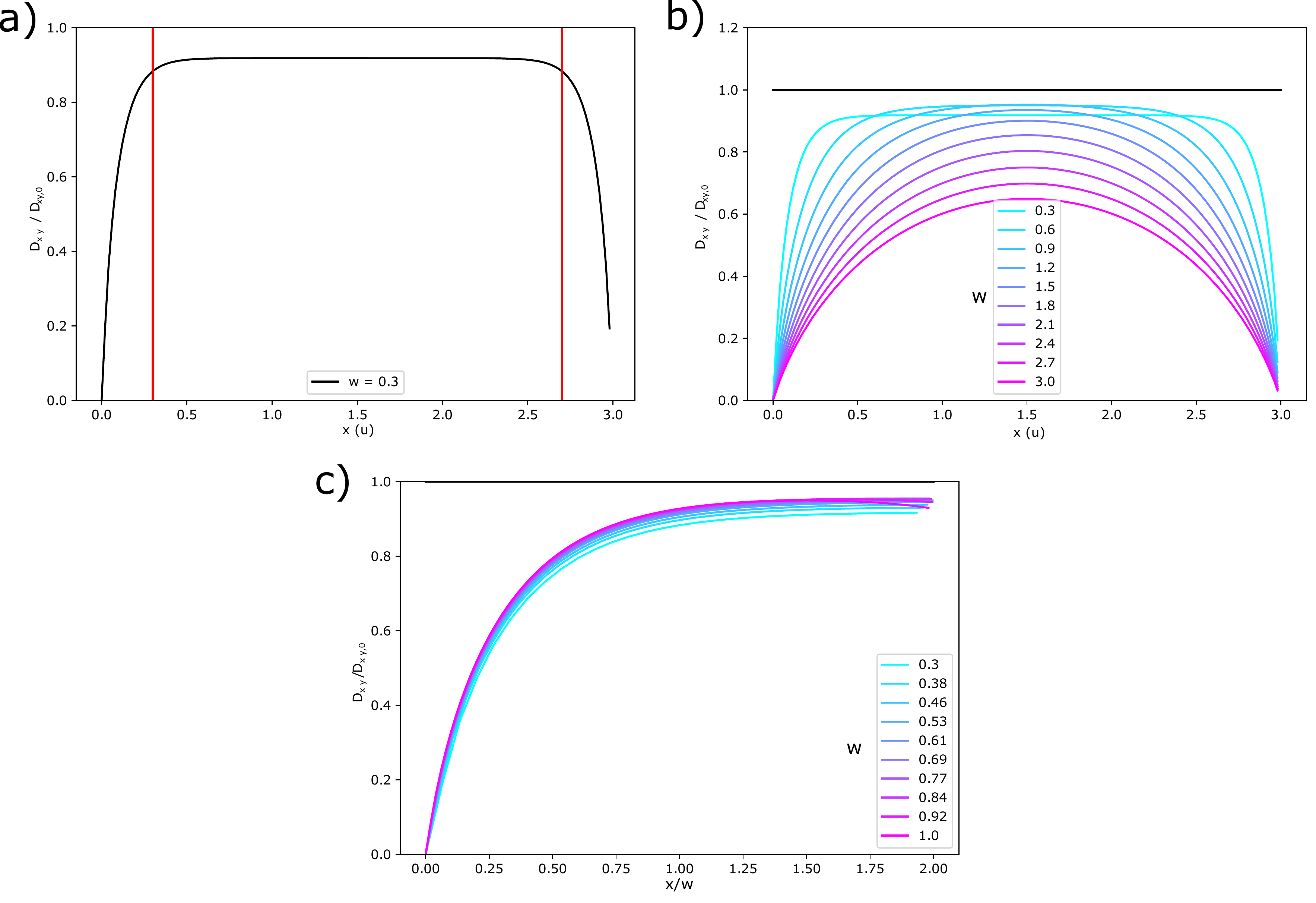}
\caption{ \label{LandW} Profile of $D_{xy}/D_{xy, 0}$ with $T=T_0$ on the bottom surface and $T = T_0+1~$K on the top surface at $x=3~$u depending on the position of transverse measurement $x$. a) Shown for $w = 0.3~$u, note that the observed value has minimal positional dependence and reasonable agreement with the true value if $\widetilde{x} > w$ from the top or bottom boundary. b) Shown for a variety of widths ranging from $w = 0.1l$ to $w = l$. The error in $D_{xy}$ measurement is $> 50~\%$ for square samples. c) Shown for a variety of $w$ as a function of $x/w$. If $w < l$ and the top and bottom boundary contributions do not interact, the error dependence is largely determined by the unitless parameter $\widetilde{x}/w$. Note that $w < 0.5l$ so that the contribution of each boundary can be isolated.}
\end{figure*}

\begin{figure*}
\centering
\includegraphics[width=\textwidth]{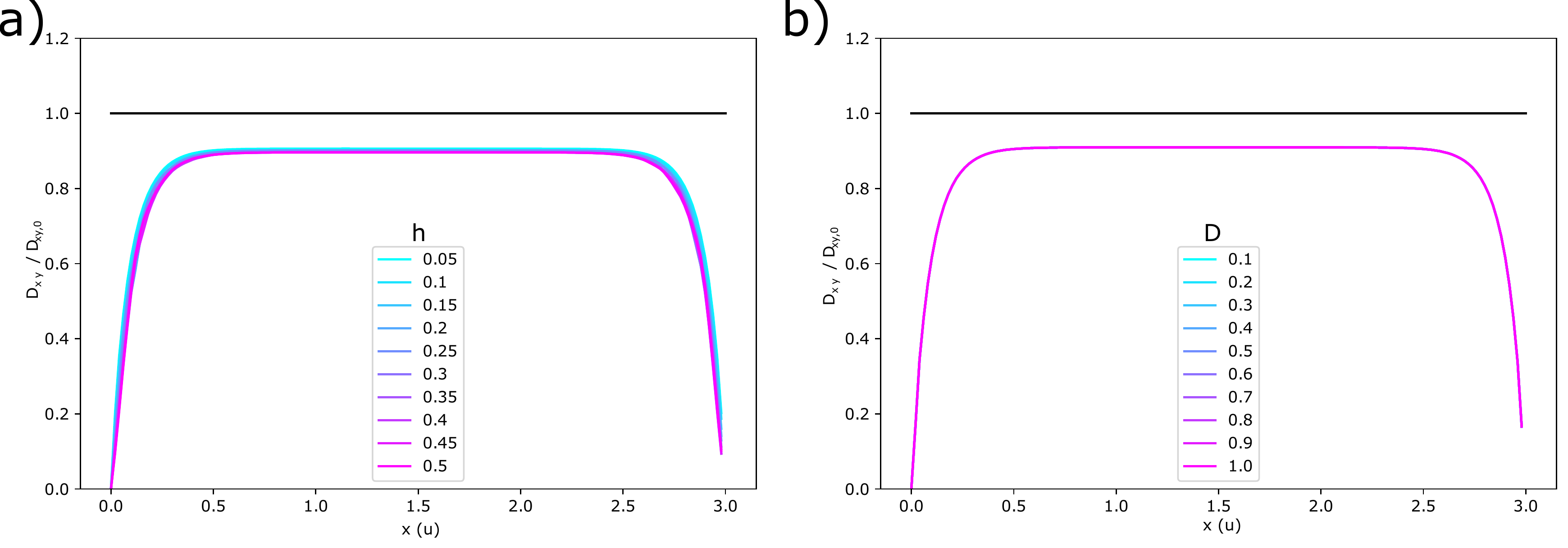}
\caption{ \label{HandD} Profile of the ratio of the observed $D_{xy}$ to $D_{xy, 0} = 0.001~$u$^{2}$/s with $T=T_0$ on the bottom surface and $T = T_0+1~$K on the top surface at $x=3~$u. a) Measured at a variety of sample heights $h$. The $h$ dependence is negligible compared to the $w$ dependence and may be attributed to slightly different simulation mesh densities with different surface area to volume ratios. b) Measured at different longitudinal thermal diffusivities $\textbf{D}$. As expected, there is no evidence of $\textbf{D}$ dependence or a hidden diffusivity-based length scale.}
\end{figure*}

The thermometer placement-based error in measured $D_{xy}$ relative to Eqn. \ref{baseEq} is determined by $w$ and the distance to the $x$ boundary, $\widetilde{x}$. As seen in Fig. \ref{LandW}, measured $D_{xy}$ is smaller than the true value and changes dramatically with contact placement if $\widetilde{x} \approx w$. Such behavior is seen for a large range of $w$ in Fig. \ref{LandW}b and the geometric correction is determined solely by the normalized distance to the boundary $\widetilde{x}/w$ as seen in in Fig. \ref{LandW}c. For square samples, $D_{xy}/D_{xy, 0}$ is 0.6 at the sample midpoint and is strongly $x$-dependent even near the sample center. In contrast, if $w~ \widetilde{<} ~0.3l$, $D_{xy}/D_{xy, 0} \sim 0.9$ at the sample midpoint and there is a large range over which observed $D_{xy}$ is relatively $x$-independent. Rectangular samples thus provide more reliable measurements of $D_{xy}$ than square samples even if non-ideal boundary conditions are imposed.

Such a simple $l$ and $w$-based geometric correction should be expected. For a sample heated with frequency $\omega$, the Green's function of Eqn. \ref{baseTrue} is a damped plane wave with wavelength $\lambda = \pi\sqrt{2D/\omega}$. This diffusivity-based length scale diverges in the DC limit, leaving only the sample dimensions as relevant length scales. As $h$ determines only the overall amplitude of $\textbf{D}$, the geometric error must be set only by $l$ and $w$. As seen in Fig. \ref{HandD}, the geometric correction factor is relatively independent of $h$ and fully independent of the longitudinal diffusivity. Such observations confirm that there are no other hidden length scales associated with the problem and that geometric corrections are determined by the sample lateral dimensions. The $\textbf{D}$ and $h$-independence of the contact placement based error simplifies the problem of simulating for a corrective geometric factor and confirms simulation accuracy.

\subsection{Mixed Boundary Conditions}

\begin{figure}
\centering
\includegraphics[width=.45\textwidth]{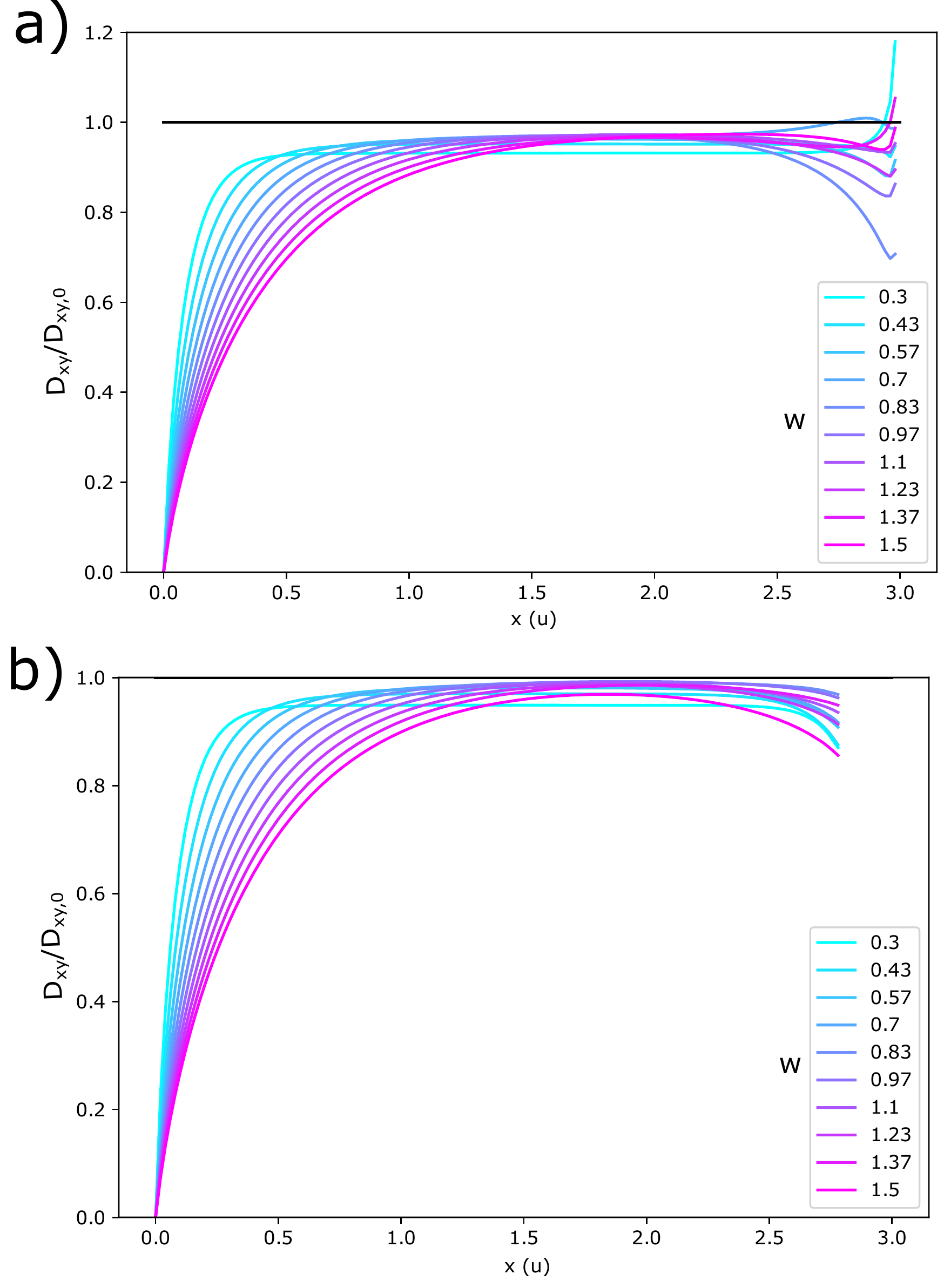}
\caption{ \label{NBC} a) Profile of the ratio of the observed $D_{yz}$ to the true value with $T=T_0$ on the bottom surface and $dT/dz = \alpha~$K/u on the top surface at $z=3~$u. The constant $\alpha$ is selected so that the maximum temperature in the sample is 1~K. b) Modeling the input heat as a heater with an arbitrary penetration depth of $0.05~$u. Note that this introduces a second length-scale to the problem, making the boundary form more complex. However, such an assumption makes simulation results more consistent between different finite element solver mesh densities.}
\end{figure}

Finally, simulations of a blended boundary condition with constant heat flow on the top surface and set temperature on the bottom surface exhibit the isolated boundary correction from one surface. Such a condition is closest to samples which are connected to a resistive heater on one end and firmly anchored to a constant temperature base on the other. As seen in Fig. \ref{NBC}a, although the observed $D_{xy}$ at $x = l$ changes due to simulation error, there is no simple systematic dependence of the top boundary $D_{xy}$ with sample size. In contrast, the length scale of the bottom boundary contribution scales with the sample width. This is observed for a variety of sample widths and two forms of sample heating in Fig. \ref{NBC}a and b.

The viable range of $w/l$ for measurement may be extended even if only one surface can exhibits uniform heat flow. Comparing the $D_{xy}$ response in Fig. \ref{NBC} and Fig. \ref{LandW}, for samples with two constant temperature surfaces, the viable measurement region must be at $\widetilde{x} > w$ from both surfaces and accurate measurement can only be performed with $w~ \widetilde{<} ~0.3l$. In contrast, with only one constant temperature surface data may be taken with $w~ \widetilde{<} ~0.5l$ provided observations are made above the sample midpoint. The boundary condition error in $D_{xy}$ measurement through a thermal Hall bar therefore can be significant but also may be accounted for easily through experimental design or simulation.

\section{Proposed Optical $D_{xy}$ Experiment}\label{MFA}

\begin{figure*}[!htb]
\centering
\includegraphics[width=\textwidth]{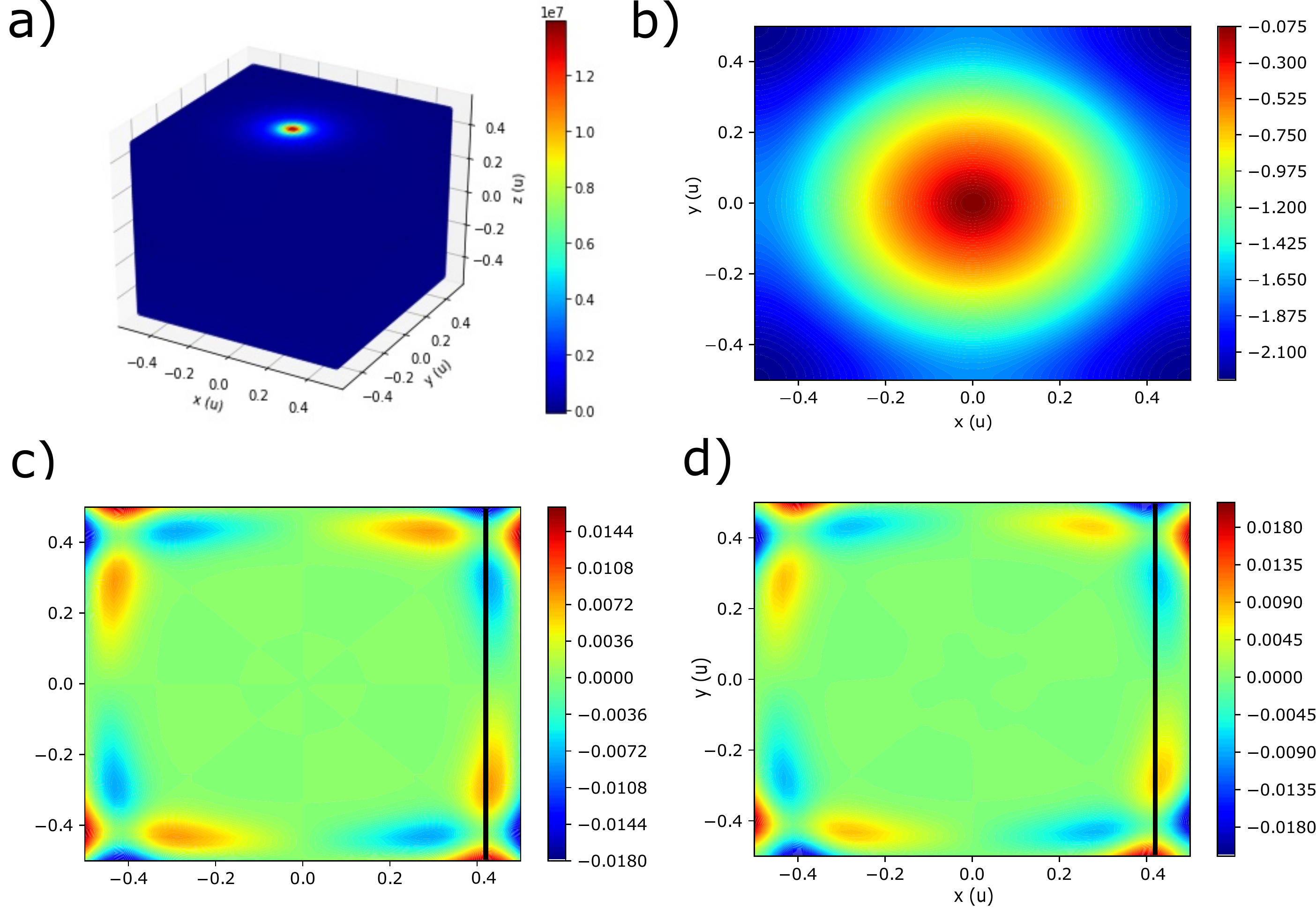}
\caption{ \label{BSM} a) Simulated real part of $T$ response to a narrow Gaussian beam on a 1x1x1 unit cube with $D_{xx} = D_{yy} = D_{zz} = 1~$u$^2$/s, $\omega = 20~$rad/s, and laser penetration depth of $0.01~$u. The $\omega$ value was chosen so $\lambda$ and the sample scale were similar. b) Simulated phase projected on the cube top surface. c) The $D_{xy}$-dependent component of the simulated phase response with $D_{xy} = 0.05$~u$^2$/s and $\omega = 200$~rad/s. d) The corresponding $D_{xy}$-dependent component of the closed form phase response model.}
\end{figure*}

A key difficulty in measuring $\kappa_{xy}$ with bulk wires and thermometers in a thermal Hall bar comes from aligning the transverse $T_{y}$ thermometers perfectly orthogonally to the applied thermal gradient. This restricts measurement to the regime where $\kappa_{xy} \approx \kappa_{xx}$ due to longitudinal pollution of the $\kappa_{xy}$ measurement. A photothermal microscope approach\cite{Zhang5378, Zhang2019} to measuring $\textbf{D}$ provides a contactless means of measuring $D_{xy}$ without longitudinal pollution. At a rescaled distance $r'$
\begin{equation}
    r' = \sqrt{\frac{x^2}{D_{xx}} + \frac{y^2}{D_{yy}} + \frac{z^2}{D_{zz}}}
\end{equation}
from an oscillatory point heat source provided by a laser there is a boundary-free temperature response
\begin{equation}
    T \propto \frac{e^{-\sqrt{i\omega}r'}}{r'}
\end{equation}
shown in Fig. \ref{BSM}a. The phase delay
\begin{equation}
    \phi(r') = \sqrt{\frac{\omega}{2}}r'
\end{equation}
can be measured with a lock-in amplifier and a second reflectivity probing laser. The phase response of a square crystal to a centered heating beam is shown in Fig. \ref{BSM}b and has been analyzed in strange metals and cuprates\cite{Zhang5378, Zhang2019}.

\begin{figure*}[!htb]
\centering
\includegraphics[width=\textwidth]{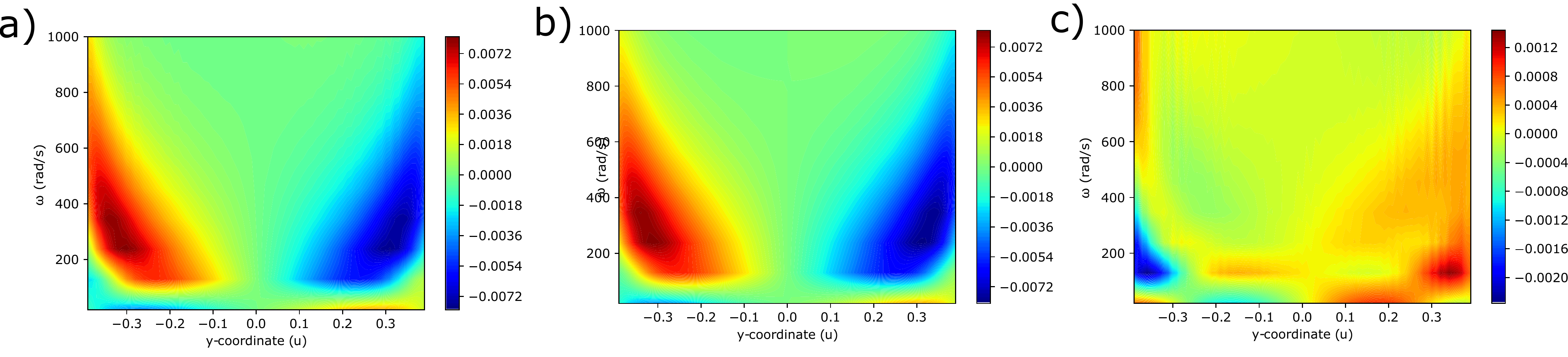}
\caption{ \label{MonopoleSweep} a) The $D_{xy}$-dependent component of the simulated phase response with $D_{xx} = D_{yy} = D_{zz} = 1$~u$^2$/s and $D_{xy} = 0.05$~u$^2$/s, sweeping $\omega$ from $20-1000$~rad/s. A linecut is taken at $x = 0.42$~u as seen in Fig. \ref{BSM}. b) The same as in subfigure a, but the closed form model result. c) Model error in the simulated solution frequency sweep. Note model error is minimal compared to the signal when data is not taken directly at the sample $y$-boundary.}
\end{figure*}

\begin{figure*}[!htb]
\centering
\includegraphics[width=\textwidth]{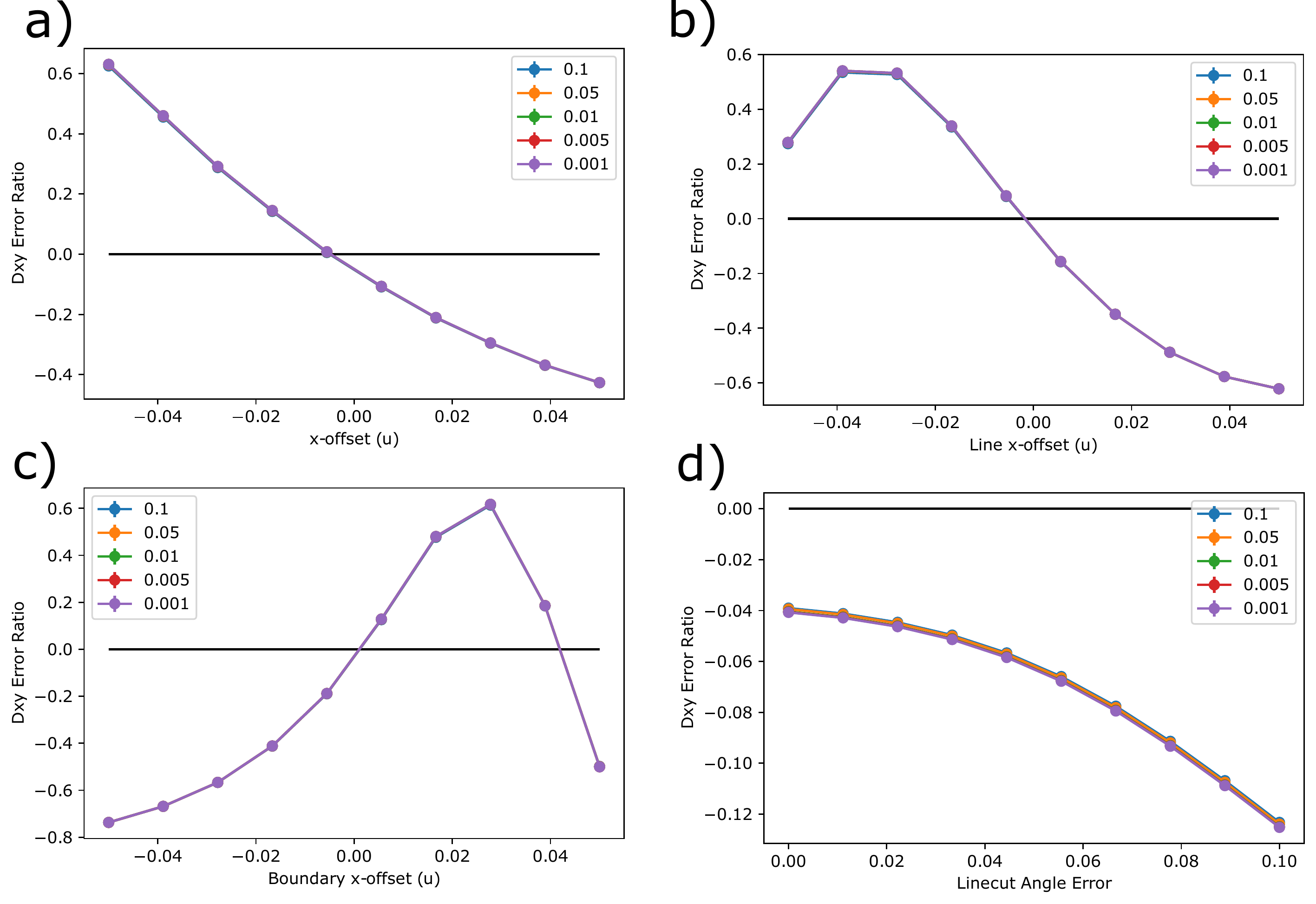}
\caption{ \label{FitError} Analyzing geometrical error for multiple $D_{xy}$ as a function of error in the measurement of a) The heating source location $x-$coordinate. b) The $x-$coordinate of the linecut of data taking. c) The location of the $x-$boundary. d) The angle of the linecut relative to the $y$-axis. Note that only the odd component of signal is used for angle analysis and that there is minimal $D_{xy}$-dependence in $Err(G')$.}
\end{figure*}

In the presence of a symmetry-breaking magnetic field, $D_{xy}$-based contributions to $\phi$ appear due near the sample boundary as seen in Fig. \ref{BSM}c. The $D_{xy}$-based $\delta \phi$ can be modeled through a series of dipole mirror sources as derived in App. \ref{Dxy} and plotted in Fig. \ref{BSM}d. This model is qualitatively indistinguishable from the simulated data for a wide range of $\omega$ and in a large region of the sample as shown in Fig. \ref{MonopoleSweep}. Finally, fitting for $D_{xy}$ using the closed form solution and simulated data yields $D_{xy} = 0.048 \pm 0.003$~u$^2$/s for a simulated $D_{xy} = 0.05$~u$^2$/s. The model error therefore is negligible compared to geometric errors in fitting for $\kappa_{xy}$ through standard thermometry.

To address the impact of geometric errors in a photo-thermal microscope measurement of $D_{xy}$, the thermal phase response can be simulated in FEniCS and then fit for $D_{xy}$ while adding geometric error in the fitting model. The model prediction is a function of $D_{xy}$ as well as $\textbf{D}$ and geometric factors $G$. In an experiment, $G$ would include the measured heating and probing beam locations as well as the sample geometry. To address misalignment or geometric errors, the phase response can be simulated in FEniCS with a set of true geometrical parameters $G$ and $D_{xy}$ in is analogous process to taking data. Observing along a line of points and changing $\omega$ creates a dataset as in Fig. \ref{MonopoleSweep}. The corresponding fitting model for $D_{xy, f}$ can be generated with a different set of parameters $G'$. This is analogous to fitting experimental data where errors have been made in the measurement of $G$. Fractional geometric error in $D_{xy}$, or $Err(G')$, is defined as
\begin{equation}
    Err(G') = \frac{D_{xy, f}(G') - D_{xy}}{D_{xy}}.
\end{equation}

The geometric error ratio is plotted in Fig. \ref{FitError} for a variety of error types. Note that $Err(G')$ is independent of $D_{xy}$. This is in contrast with a thermal Hall bar, in which longitudinal pollution creates an offset $\delta \kappa_{xy}$ that depends only on $\kappa_{xx}$ and thus diverges as $\kappa_{xy} \rightarrow 0$. The contactless thermal microscope technique therefore does not exhibit the misalignment pollution of Hall bar techniques. Additionally, the spacial unit $u$ is defined in reference to the sample size and corresponding frequency-controlled heat diffusion wavelength. The largest geometric parameter uncertainty is the location of the heating and probing beams due to their widths of $\sim 1~\mu$m. The relevant $x$-offset in Fig. \ref{FitError} for a sample of length scale $l$ therefore is $1~\mu$m/$l$. Geometric errors correspondingly can be ignored for samples larger than 1$~$mm where such errors are smaller than the model error.

\section{Summary}

The above discussion yields important guidelines for the measurement of thermal Hall coefficient, especially when the amplitude of the effect is important. The geometric corrections can be summarized as follows:
\begin{itemize}
    \item Sample geometry and contact placement can significantly effect the measured $D_{xy}$ or $\kappa_{xy}$ in laterally square samples. As such an error is $D_{xx}$-independent, it would not be accounted for through antisymmetrization with magnetic field and is difficult to correct for.
    \item Rectangular samples with $w < 0.3l$ should produce more consistent measurements of transverse thermal properties regardless of the thermal anchoring method. This condition is commonly met in $\kappa_{xy}$ measurement where the amplitude of the signal is important and sample dimensions are listed\cite{Ong, kasahara_2018,kasahara_PRL, Behnia_2017, Behnia_2020, Wyder, kim_2019}.
    \item A geometric corrective factor can be simulated easily once per sample and used to increase confidence in the reproducibility of transverse thermal transport properties between samples if boundary conditions are known.
\end{itemize}

In conclusion, as theories become more detailed with material parameters for comparison to experiments, it is important that experimental data is stated accounting for possible sample geometric error. Unlike electrical Hall bars with patterned contacts and current sources, there are a variety of methods used to control heat flow and temperature in thermal Hall bars. As boundary conditions can create large corrective factors for square samples, the chosen contact method and sample shapes are important elements in thermal Hall measurement. Finally, with the increased interest in transverse thermal property measurement and the large misalignment signal caused by bulky thermometers, a contactless optical method for measuring $D_{xy}$ could provide greater accuracy in future studies.

\section*{Acknowledgements}

This work was funded by the Army Research Office grant W911NF1710588, and by the Gordon and by the Department of Energy, Office of Basic Energy Sciences, under contract no. DE-AC02-76SF00515.

\section*{Data Availability}

The AIP is committed to data availability so readers can replicate, interpret, and build upon the work in each paper. The data that support the findings of this study are available from the corresponding author upon reasonable request. All code used to generate simulated Hall Bar data is commented and included in github at DOI reference number 10.5281/zenodo.3974941 and the code used for optical simulation is included at DOI 10.5281/zenodo.3974945.

\bibliography{bibl.bib}

\begin{thebibliography}{19}%
\makeatletter
\providecommand \@ifxundefined [1]{%
 \@ifx{#1\undefined}
}%
\providecommand \@ifnum [1]{%
 \ifnum #1\expandafter \@firstoftwo
 \else \expandafter \@secondoftwo
 \fi
}%
\providecommand \@ifx [1]{%
 \ifx #1\expandafter \@firstoftwo
 \else \expandafter \@secondoftwo
 \fi
}%
\providecommand \natexlab [1]{#1}%
\providecommand \enquote  [1]{``#1''}%
\providecommand \bibnamefont  [1]{#1}%
\providecommand \bibfnamefont [1]{#1}%
\providecommand \citenamefont [1]{#1}%
\providecommand \href@noop [0]{\@secondoftwo}%
\providecommand \href [0]{\begingroup \@sanitize@url \@href}%
\providecommand \@href[1]{\@@startlink{#1}\@@href}%
\providecommand \@@href[1]{\endgroup#1\@@endlink}%
\providecommand \@sanitize@url [0]{\catcode `\\12\catcode `\$12\catcode
  `\&12\catcode `\#12\catcode `\^12\catcode `\_12\catcode `\%12\relax}%
\providecommand \@@startlink[1]{}%
\providecommand \@@endlink[0]{}%
\providecommand \url  [0]{\begingroup\@sanitize@url \@url }%
\providecommand \@url [1]{\endgroup\@href {#1}{\urlprefix }}%
\providecommand \urlprefix  [0]{URL }%
\providecommand \Eprint [0]{\href }%
\providecommand \doibase [0]{http://dx.doi.org/}%
\providecommand \selectlanguage [0]{\@gobble}%
\providecommand \bibinfo  [0]{\@secondoftwo}%
\providecommand \bibfield  [0]{\@secondoftwo}%
\providecommand \translation [1]{[#1]}%
\providecommand \BibitemOpen [0]{}%
\providecommand \bibitemStop [0]{}%
\providecommand \bibitemNoStop [0]{.\EOS\space}%
\providecommand \EOS [0]{\spacefactor3000\relax}%
\providecommand \BibitemShut  [1]{\csname bibitem#1\endcsname}%
\let\auto@bib@innerbib\@empty
\bibitem [{\citenamefont {Hirschberger}\ \emph
  {et~al.}(2015{\natexlab{a}})\citenamefont {Hirschberger}, \citenamefont
  {Krizan}, \citenamefont {Cava},\ and\ \citenamefont {Ong}}]{Ong}%
  \BibitemOpen
  \bibfield  {author} {\bibinfo {author} {\bibfnamefont {M.}~\bibnamefont
  {Hirschberger}}, \bibinfo {author} {\bibfnamefont {J.~W.}\ \bibnamefont
  {Krizan}}, \bibinfo {author} {\bibfnamefont {R.~J.}\ \bibnamefont {Cava}}, \
  and\ \bibinfo {author} {\bibfnamefont {N.~P.}\ \bibnamefont {Ong}},\
  }\bibfield  {title} {\enquote {\bibinfo {title} {Large thermal hall
  conductivity of neutral spin excitations in a frustrated quantum magnet},}\
  }\href {\doibase 10.1126/science.1257340} {\bibfield  {journal} {\bibinfo
  {journal} {Science}\ }\textbf {\bibinfo {volume} {348}},\ \bibinfo {pages}
  {106--109} (\bibinfo {year} {2015}{\natexlab{a}})},\ \Eprint
  {http://arxiv.org/abs/https://science.sciencemag.org/content/348/6230/106.full.pdf}
  {https://science.sciencemag.org/content/348/6230/106.full.pdf} \BibitemShut
  {NoStop}%
\bibitem [{\citenamefont {Sugii}\ \emph {et~al.}(2017)\citenamefont {Sugii},
  \citenamefont {Shimozawa}, \citenamefont {Watanabe}, \citenamefont {Suzuki},
  \citenamefont {Halim}, \citenamefont {Kimata}, \citenamefont {Matsumoto},
  \citenamefont {Nakatsuji},\ and\ \citenamefont {Yamashita}}]{Yamashita}%
  \BibitemOpen
  \bibfield  {author} {\bibinfo {author} {\bibfnamefont {K.}~\bibnamefont
  {Sugii}}, \bibinfo {author} {\bibfnamefont {M.}~\bibnamefont {Shimozawa}},
  \bibinfo {author} {\bibfnamefont {D.}~\bibnamefont {Watanabe}}, \bibinfo
  {author} {\bibfnamefont {Y.}~\bibnamefont {Suzuki}}, \bibinfo {author}
  {\bibfnamefont {M.}~\bibnamefont {Halim}}, \bibinfo {author} {\bibfnamefont
  {M.}~\bibnamefont {Kimata}}, \bibinfo {author} {\bibfnamefont
  {Y.}~\bibnamefont {Matsumoto}}, \bibinfo {author} {\bibfnamefont
  {S.}~\bibnamefont {Nakatsuji}}, \ and\ \bibinfo {author} {\bibfnamefont
  {M.}~\bibnamefont {Yamashita}},\ }\bibfield  {title} {\enquote {\bibinfo
  {title} {Thermal hall effect in a phonon-glass
  ${\mathrm{ba}}_{3}{\mathrm{cusb}}_{2}{\mathrm{o}}_{9}$},}\ }\href {\doibase
  10.1103/PhysRevLett.118.145902} {\bibfield  {journal} {\bibinfo  {journal}
  {Phys. Rev. Lett.}\ }\textbf {\bibinfo {volume} {118}},\ \bibinfo {pages}
  {145902} (\bibinfo {year} {2017})}\BibitemShut {NoStop}%
\bibitem [{\citenamefont {Watanabe}\ \emph {et~al.}(2016)\citenamefont
  {Watanabe}, \citenamefont {Sugii}, \citenamefont {Shimozawa}, \citenamefont
  {Suzuki}, \citenamefont {Yajima}, \citenamefont {Ishikawa}, \citenamefont
  {Hiroi}, \citenamefont {Shibauchi}, \citenamefont {Matsuda},\ and\
  \citenamefont {Yamashita}}]{Watanabe2016}%
  \BibitemOpen
  \bibfield  {author} {\bibinfo {author} {\bibfnamefont {D.}~\bibnamefont
  {Watanabe}}, \bibinfo {author} {\bibfnamefont {K.}~\bibnamefont {Sugii}},
  \bibinfo {author} {\bibfnamefont {M.}~\bibnamefont {Shimozawa}}, \bibinfo
  {author} {\bibfnamefont {Y.}~\bibnamefont {Suzuki}}, \bibinfo {author}
  {\bibfnamefont {T.}~\bibnamefont {Yajima}}, \bibinfo {author} {\bibfnamefont
  {H.}~\bibnamefont {Ishikawa}}, \bibinfo {author} {\bibfnamefont
  {Z.}~\bibnamefont {Hiroi}}, \bibinfo {author} {\bibfnamefont
  {T.}~\bibnamefont {Shibauchi}}, \bibinfo {author} {\bibfnamefont
  {Y.}~\bibnamefont {Matsuda}}, \ and\ \bibinfo {author} {\bibfnamefont
  {M.}~\bibnamefont {Yamashita}},\ }\bibfield  {title} {\enquote {\bibinfo
  {title} {Emergence of nontrivial magnetic excitations in a spin-liquid state
  of kagom{\'e} volborthite},}\ }\href {\doibase 10.1073/pnas.1524076113}
  {\bibfield  {journal} {\bibinfo  {journal} {Proceedings of the National
  Academy of Sciences}\ }\textbf {\bibinfo {volume} {113}},\ \bibinfo {pages}
  {8653--8657} (\bibinfo {year} {2016})},\ \Eprint
  {http://arxiv.org/abs/https://www.pnas.org/content/113/31/8653.full.pdf}
  {https://www.pnas.org/content/113/31/8653.full.pdf} \BibitemShut {NoStop}%
\bibitem [{\citenamefont {Kasahara}\ \emph
  {et~al.}(2018{\natexlab{a}})\citenamefont {Kasahara}, \citenamefont
  {Ohnishi}, \citenamefont {Mizukami}, \citenamefont {Tanaka}, \citenamefont
  {Ma}, \citenamefont {Sugii}, \citenamefont {Kurita}, \citenamefont {Tanaka},
  \citenamefont {Nasu}, \citenamefont {Motome},\ and\ \citenamefont
  {et~al.}}]{kasahara_2018}%
  \BibitemOpen
  \bibfield  {author} {\bibinfo {author} {\bibfnamefont {Y.}~\bibnamefont
  {Kasahara}}, \bibinfo {author} {\bibfnamefont {T.}~\bibnamefont {Ohnishi}},
  \bibinfo {author} {\bibfnamefont {Y.}~\bibnamefont {Mizukami}}, \bibinfo
  {author} {\bibfnamefont {O.}~\bibnamefont {Tanaka}}, \bibinfo {author}
  {\bibfnamefont {S.}~\bibnamefont {Ma}}, \bibinfo {author} {\bibfnamefont
  {K.}~\bibnamefont {Sugii}}, \bibinfo {author} {\bibfnamefont
  {N.}~\bibnamefont {Kurita}}, \bibinfo {author} {\bibfnamefont
  {H.}~\bibnamefont {Tanaka}}, \bibinfo {author} {\bibfnamefont
  {J.}~\bibnamefont {Nasu}}, \bibinfo {author} {\bibfnamefont {Y.}~\bibnamefont
  {Motome}}, \ and\ \bibinfo {author} {\bibnamefont {et~al.}},\ }\bibfield
  {title} {\enquote {\bibinfo {title} {Majorana quantization and half-integer
  thermal quantum hall effect in a kitaev spin liquid},}\ }\href {\doibase
  10.1038/s41586-018-0274-0} {\bibfield  {journal} {\bibinfo  {journal}
  {Nature}\ }\textbf {\bibinfo {volume} {559}},\ \bibinfo {pages} {227–231}
  (\bibinfo {year} {2018}{\natexlab{a}})}\BibitemShut {NoStop}%
\bibitem [{\citenamefont {Kasahara}\ \emph
  {et~al.}(2018{\natexlab{b}})\citenamefont {Kasahara}, \citenamefont {Sugii},
  \citenamefont {Ohnishi}, \citenamefont {Shimozawa}, \citenamefont
  {Yamashita}, \citenamefont {Kurita}, \citenamefont {Tanaka}, \citenamefont
  {Nasu}, \citenamefont {Motome}, \citenamefont {Shibauchi},\ and\
  \citenamefont {et~al.}}]{kasahara_PRL}%
  \BibitemOpen
  \bibfield  {author} {\bibinfo {author} {\bibfnamefont {Y.}~\bibnamefont
  {Kasahara}}, \bibinfo {author} {\bibfnamefont {K.}~\bibnamefont {Sugii}},
  \bibinfo {author} {\bibfnamefont {T.}~\bibnamefont {Ohnishi}}, \bibinfo
  {author} {\bibfnamefont {M.}~\bibnamefont {Shimozawa}}, \bibinfo {author}
  {\bibfnamefont {M.}~\bibnamefont {Yamashita}}, \bibinfo {author}
  {\bibfnamefont {N.}~\bibnamefont {Kurita}}, \bibinfo {author} {\bibfnamefont
  {H.}~\bibnamefont {Tanaka}}, \bibinfo {author} {\bibfnamefont
  {J.}~\bibnamefont {Nasu}}, \bibinfo {author} {\bibfnamefont {Y.}~\bibnamefont
  {Motome}}, \bibinfo {author} {\bibfnamefont {T.}~\bibnamefont {Shibauchi}}, \
  and\ \bibinfo {author} {\bibnamefont {et~al.}},\ }\bibfield  {title}
  {\enquote {\bibinfo {title} {Unusual thermal hall effect in a kitaev spin
  liquid candidate rucl3},}\ }\href {\doibase 10.1103/physrevlett.120.217205}
  {\bibfield  {journal} {\bibinfo  {journal} {Physical Review Letters}\
  }\textbf {\bibinfo {volume} {120}} (\bibinfo {year} {2018}{\natexlab{b}}),\
  10.1103/physrevlett.120.217205}\BibitemShut {NoStop}%
\bibitem [{\citenamefont {Doki}\ \emph {et~al.}(2018)\citenamefont {Doki},
  \citenamefont {Akazawa}, \citenamefont {Lee}, \citenamefont {Han},
  \citenamefont {Sugii}, \citenamefont {Shimozawa}, \citenamefont {Kawashima},
  \citenamefont {Oda}, \citenamefont {Yoshida},\ and\ \citenamefont
  {Yamashita}}]{Doki_2018}%
  \BibitemOpen
  \bibfield  {author} {\bibinfo {author} {\bibfnamefont {H.}~\bibnamefont
  {Doki}}, \bibinfo {author} {\bibfnamefont {M.}~\bibnamefont {Akazawa}},
  \bibinfo {author} {\bibfnamefont {H.-Y.}\ \bibnamefont {Lee}}, \bibinfo
  {author} {\bibfnamefont {J.~H.}\ \bibnamefont {Han}}, \bibinfo {author}
  {\bibfnamefont {K.}~\bibnamefont {Sugii}}, \bibinfo {author} {\bibfnamefont
  {M.}~\bibnamefont {Shimozawa}}, \bibinfo {author} {\bibfnamefont
  {N.}~\bibnamefont {Kawashima}}, \bibinfo {author} {\bibfnamefont
  {M.}~\bibnamefont {Oda}}, \bibinfo {author} {\bibfnamefont {H.}~\bibnamefont
  {Yoshida}}, \ and\ \bibinfo {author} {\bibfnamefont {M.}~\bibnamefont
  {Yamashita}},\ }\bibfield  {title} {\enquote {\bibinfo {title} {Spin thermal
  hall conductivity of a kagome antiferromagnet},}\ }\href {\doibase
  10.1103/PhysRevLett.121.097203} {\bibfield  {journal} {\bibinfo  {journal}
  {Phys. Rev. Lett.}\ }\textbf {\bibinfo {volume} {121}},\ \bibinfo {pages}
  {097203} (\bibinfo {year} {2018})}\BibitemShut {NoStop}%
\bibitem [{\citenamefont {Hentrich}\ \emph {et~al.}(2019)\citenamefont
  {Hentrich}, \citenamefont {Roslova}, \citenamefont {Isaeva}, \citenamefont
  {Doert}, \citenamefont {Brenig}, \citenamefont {B\"uchner},\ and\
  \citenamefont {Hess}}]{Hentrich_2019}%
  \BibitemOpen
  \bibfield  {author} {\bibinfo {author} {\bibfnamefont {R.}~\bibnamefont
  {Hentrich}}, \bibinfo {author} {\bibfnamefont {M.}~\bibnamefont {Roslova}},
  \bibinfo {author} {\bibfnamefont {A.}~\bibnamefont {Isaeva}}, \bibinfo
  {author} {\bibfnamefont {T.}~\bibnamefont {Doert}}, \bibinfo {author}
  {\bibfnamefont {W.}~\bibnamefont {Brenig}}, \bibinfo {author} {\bibfnamefont
  {B.}~\bibnamefont {B\"uchner}}, \ and\ \bibinfo {author} {\bibfnamefont
  {C.}~\bibnamefont {Hess}},\ }\bibfield  {title} {\enquote {\bibinfo {title}
  {Large thermal hall effect in
  $\ensuremath{\alpha}\text{\ensuremath{-}}{\mathrm{rucl}}_{3}$: Evidence for
  heat transport by kitaev-heisenberg paramagnons},}\ }\href {\doibase
  10.1103/PhysRevB.99.085136} {\bibfield  {journal} {\bibinfo  {journal} {Phys.
  Rev. B}\ }\textbf {\bibinfo {volume} {99}},\ \bibinfo {pages} {085136}
  (\bibinfo {year} {2019})}\BibitemShut {NoStop}%
\bibitem [{\citenamefont {Hirschberger}\ \emph
  {et~al.}(2015{\natexlab{b}})\citenamefont {Hirschberger}, \citenamefont
  {Chisnell}, \citenamefont {Lee},\ and\ \citenamefont
  {Ong}}]{Hirschberger2015}%
  \BibitemOpen
  \bibfield  {author} {\bibinfo {author} {\bibfnamefont {M.}~\bibnamefont
  {Hirschberger}}, \bibinfo {author} {\bibfnamefont {R.}~\bibnamefont
  {Chisnell}}, \bibinfo {author} {\bibfnamefont {Y.~S.}\ \bibnamefont {Lee}}, \
  and\ \bibinfo {author} {\bibfnamefont {N.~P.}\ \bibnamefont {Ong}},\
  }\bibfield  {title} {\enquote {\bibinfo {title} {Thermal hall effect of spin
  excitations in a kagome magnet},}\ }\href {\doibase
  10.1103/PhysRevLett.115.106603} {\bibfield  {journal} {\bibinfo  {journal}
  {Phys. Rev. Lett.}\ }\textbf {\bibinfo {volume} {115}},\ \bibinfo {pages}
  {106603} (\bibinfo {year} {2015}{\natexlab{b}})}\BibitemShut {NoStop}%
\bibitem [{\citenamefont {Ideue}\ \emph {et~al.}(2012)\citenamefont {Ideue},
  \citenamefont {Onose}, \citenamefont {Katsura}, \citenamefont {Shiomi},
  \citenamefont {Ishiwata}, \citenamefont {Nagaosa},\ and\ \citenamefont
  {Tokura}}]{Ideue2012}%
  \BibitemOpen
  \bibfield  {author} {\bibinfo {author} {\bibfnamefont {T.}~\bibnamefont
  {Ideue}}, \bibinfo {author} {\bibfnamefont {Y.}~\bibnamefont {Onose}},
  \bibinfo {author} {\bibfnamefont {H.}~\bibnamefont {Katsura}}, \bibinfo
  {author} {\bibfnamefont {Y.}~\bibnamefont {Shiomi}}, \bibinfo {author}
  {\bibfnamefont {S.}~\bibnamefont {Ishiwata}}, \bibinfo {author}
  {\bibfnamefont {N.}~\bibnamefont {Nagaosa}}, \ and\ \bibinfo {author}
  {\bibfnamefont {Y.}~\bibnamefont {Tokura}},\ }\bibfield  {title} {\enquote
  {\bibinfo {title} {Effect of lattice geometry on magnon hall effect in
  ferromagnetic insulators},}\ }\href {\doibase 10.1103/PhysRevB.85.134411}
  {\bibfield  {journal} {\bibinfo  {journal} {Phys. Rev. B}\ }\textbf {\bibinfo
  {volume} {85}},\ \bibinfo {pages} {134411} (\bibinfo {year}
  {2012})}\BibitemShut {NoStop}%
\bibitem [{\citenamefont {Onose}\ \emph {et~al.}(2010)\citenamefont {Onose},
  \citenamefont {Ideue}, \citenamefont {Katsura}, \citenamefont {Shiomi},
  \citenamefont {Nagaosa},\ and\ \citenamefont {Tokura}}]{Onose2010}%
  \BibitemOpen
  \bibfield  {author} {\bibinfo {author} {\bibfnamefont {Y.}~\bibnamefont
  {Onose}}, \bibinfo {author} {\bibfnamefont {T.}~\bibnamefont {Ideue}},
  \bibinfo {author} {\bibfnamefont {H.}~\bibnamefont {Katsura}}, \bibinfo
  {author} {\bibfnamefont {Y.}~\bibnamefont {Shiomi}}, \bibinfo {author}
  {\bibfnamefont {N.}~\bibnamefont {Nagaosa}}, \ and\ \bibinfo {author}
  {\bibfnamefont {Y.}~\bibnamefont {Tokura}},\ }\bibfield  {title} {\enquote
  {\bibinfo {title} {Observation of the magnon hall effect},}\ }\href {\doibase
  10.1126/science.1188260} {\bibfield  {journal} {\bibinfo  {journal}
  {Science}\ }\textbf {\bibinfo {volume} {329}},\ \bibinfo {pages} {297--299}
  (\bibinfo {year} {2010})},\ \Eprint
  {http://arxiv.org/abs/https://science.sciencemag.org/content/329/5989/297.full.pdf}
  {https://science.sciencemag.org/content/329/5989/297.full.pdf} \BibitemShut
  {NoStop}%
\bibitem [{\citenamefont {Inyushkin}\ and\ \citenamefont
  {Taldenkov}(2007)}]{Taldenkov}%
  \BibitemOpen
  \bibfield  {author} {\bibinfo {author} {\bibfnamefont {A.~V.}\ \bibnamefont
  {Inyushkin}}\ and\ \bibinfo {author} {\bibfnamefont {A.~N.}\ \bibnamefont
  {Taldenkov}},\ }\bibfield  {title} {\enquote {\bibinfo {title} {On the phonon
  hall effect in a paramagnetic dielectric},}\ }\href {\doibase
  10.1134/s0021364007180075} {\bibfield  {journal} {\bibinfo  {journal} {JETP
  Letters}\ }\textbf {\bibinfo {volume} {86}},\ \bibinfo {pages} {379–382}
  (\bibinfo {year} {2007})}\BibitemShut {NoStop}%
\bibitem [{\citenamefont {Strohm}, \citenamefont {Rikken},\ and\ \citenamefont
  {Wyder}(2005)}]{Wyder}%
  \BibitemOpen
  \bibfield  {author} {\bibinfo {author} {\bibfnamefont {C.}~\bibnamefont
  {Strohm}}, \bibinfo {author} {\bibfnamefont {G.~L. J.~A.}\ \bibnamefont
  {Rikken}}, \ and\ \bibinfo {author} {\bibfnamefont {P.}~\bibnamefont
  {Wyder}},\ }\bibfield  {title} {\enquote {\bibinfo {title} {Phenomenological
  evidence for the phonon hall effect},}\ }\href {\doibase
  10.1103/PhysRevLett.95.155901} {\bibfield  {journal} {\bibinfo  {journal}
  {Phys. Rev. Lett.}\ }\textbf {\bibinfo {volume} {95}},\ \bibinfo {pages}
  {155901} (\bibinfo {year} {2005})}\BibitemShut {NoStop}%
\bibitem [{\citenamefont {Li}\ \emph {et~al.}(2020)\citenamefont {Li},
  \citenamefont {Fauqu\'e}, \citenamefont {Zhu},\ and\ \citenamefont
  {Behnia}}]{Behnia2020}%
  \BibitemOpen
  \bibfield  {author} {\bibinfo {author} {\bibfnamefont {X.}~\bibnamefont
  {Li}}, \bibinfo {author} {\bibfnamefont {B.}~\bibnamefont {Fauqu\'e}},
  \bibinfo {author} {\bibfnamefont {Z.}~\bibnamefont {Zhu}}, \ and\ \bibinfo
  {author} {\bibfnamefont {K.}~\bibnamefont {Behnia}},\ }\bibfield  {title}
  {\enquote {\bibinfo {title} {Phonon thermal hall effect in strontium
  titanate},}\ }\href {\doibase 10.1103/PhysRevLett.124.105901} {\bibfield
  {journal} {\bibinfo  {journal} {Phys. Rev. Lett.}\ }\textbf {\bibinfo
  {volume} {124}},\ \bibinfo {pages} {105901} (\bibinfo {year}
  {2020})}\BibitemShut {NoStop}%
\bibitem [{\citenamefont {Aln{\ae}s}\ \emph {et~al.}(2015)\citenamefont
  {Aln{\ae}s}, \citenamefont {Blechta}, \citenamefont {Hake}, \citenamefont
  {Johansson}, \citenamefont {Kehlet}, \citenamefont {Logg}, \citenamefont
  {Richardson}, \citenamefont {Ring}, \citenamefont {Rognes},\ and\
  \citenamefont {Wells}}]{fenics}%
  \BibitemOpen
  \bibfield  {author} {\bibinfo {author} {\bibfnamefont {M.~S.}\ \bibnamefont
  {Aln{\ae}s}}, \bibinfo {author} {\bibfnamefont {J.}~\bibnamefont {Blechta}},
  \bibinfo {author} {\bibfnamefont {J.}~\bibnamefont {Hake}}, \bibinfo {author}
  {\bibfnamefont {A.}~\bibnamefont {Johansson}}, \bibinfo {author}
  {\bibfnamefont {B.}~\bibnamefont {Kehlet}}, \bibinfo {author} {\bibfnamefont
  {A.}~\bibnamefont {Logg}}, \bibinfo {author} {\bibfnamefont {C.}~\bibnamefont
  {Richardson}}, \bibinfo {author} {\bibfnamefont {J.}~\bibnamefont {Ring}},
  \bibinfo {author} {\bibfnamefont {M.~E.}\ \bibnamefont {Rognes}}, \ and\
  \bibinfo {author} {\bibfnamefont {G.~N.}\ \bibnamefont {Wells}},\ }\bibfield
  {title} {\enquote {\bibinfo {title} {The fenics project version 1.5},}\
  }\href {\doibase 10.11588/ans.2015.100.20553} {\bibfield  {journal} {\bibinfo
   {journal} {Archive of Numerical Software}\ }\textbf {\bibinfo {volume} {3}}
  (\bibinfo {year} {2015}),\ 10.11588/ans.2015.100.20553}\BibitemShut {NoStop}%
\bibitem [{\citenamefont {Logg}, \citenamefont {Wells},\ and\ \citenamefont
  {Hake}(2012)}]{dolfin}%
  \BibitemOpen
  \bibfield  {author} {\bibinfo {author} {\bibfnamefont {A.}~\bibnamefont
  {Logg}}, \bibinfo {author} {\bibfnamefont {G.~N.}\ \bibnamefont {Wells}}, \
  and\ \bibinfo {author} {\bibfnamefont {J.}~\bibnamefont {Hake}},\ }\enquote
  {\bibinfo {title} {Dolfin: a c++/python finite element library},}\ in\
  \href@noop {} {\emph {\bibinfo {booktitle} {Automated Solution of
  Differential Equations by the Finite Element Method, Volume 84 of Lecture
  Notes in Computational Science and Engineering}}},\ \bibinfo {editor} {edited
  by\ \bibinfo {editor} {\bibfnamefont {A.}~\bibnamefont {Logg}}, \bibinfo
  {editor} {\bibfnamefont {K.-A.}\ \bibnamefont {Mardal}}, \ and\ \bibinfo
  {editor} {\bibfnamefont {G.~N.}\ \bibnamefont {Wells}}}\ (\bibinfo
  {publisher} {Springer},\ \bibinfo {year} {2012})\ Chap.~\bibinfo {chapter}
  {10}\BibitemShut {NoStop}%
\bibitem [{\citenamefont {Grissonnanche}\ \emph {et~al.}(2016)\citenamefont
  {Grissonnanche}, \citenamefont {Lalibert\'e}, \citenamefont
  {Dufour-Beaus\'ejour}, \citenamefont {Matusiak}, \citenamefont {Badoux},
  \citenamefont {Tafti}, \citenamefont {Michon}, \citenamefont {Riopel},
  \citenamefont {Cyr-Choini\`ere}, \citenamefont {Baglo}, \citenamefont
  {Ramshaw}, \citenamefont {Liang}, \citenamefont {Bonn}, \citenamefont
  {Hardy}, \citenamefont {Kr\"amer}, \citenamefont {LeBoeuf}, \citenamefont
  {Graf}, \citenamefont {Doiron-Leyraud},\ and\ \citenamefont
  {Taillefer}}]{Taillefer2016}%
  \BibitemOpen
  \bibfield  {author} {\bibinfo {author} {\bibfnamefont {G.}~\bibnamefont
  {Grissonnanche}}, \bibinfo {author} {\bibfnamefont {F.}~\bibnamefont
  {Lalibert\'e}}, \bibinfo {author} {\bibfnamefont {S.}~\bibnamefont
  {Dufour-Beaus\'ejour}}, \bibinfo {author} {\bibfnamefont {M.}~\bibnamefont
  {Matusiak}}, \bibinfo {author} {\bibfnamefont {S.}~\bibnamefont {Badoux}},
  \bibinfo {author} {\bibfnamefont {F.~F.}\ \bibnamefont {Tafti}}, \bibinfo
  {author} {\bibfnamefont {B.}~\bibnamefont {Michon}}, \bibinfo {author}
  {\bibfnamefont {A.}~\bibnamefont {Riopel}}, \bibinfo {author} {\bibfnamefont
  {O.}~\bibnamefont {Cyr-Choini\`ere}}, \bibinfo {author} {\bibfnamefont
  {J.~C.}\ \bibnamefont {Baglo}}, \bibinfo {author} {\bibfnamefont {B.~J.}\
  \bibnamefont {Ramshaw}}, \bibinfo {author} {\bibfnamefont {R.}~\bibnamefont
  {Liang}}, \bibinfo {author} {\bibfnamefont {D.~A.}\ \bibnamefont {Bonn}},
  \bibinfo {author} {\bibfnamefont {W.~N.}\ \bibnamefont {Hardy}}, \bibinfo
  {author} {\bibfnamefont {S.}~\bibnamefont {Kr\"amer}}, \bibinfo {author}
  {\bibfnamefont {D.}~\bibnamefont {LeBoeuf}}, \bibinfo {author} {\bibfnamefont
  {D.}~\bibnamefont {Graf}}, \bibinfo {author} {\bibfnamefont {N.}~\bibnamefont
  {Doiron-Leyraud}}, \ and\ \bibinfo {author} {\bibfnamefont {L.}~\bibnamefont
  {Taillefer}},\ }\bibfield  {title} {\enquote {\bibinfo {title}
  {Wiedemann-franz law in the underdoped cuprate superconductor
  ${\mathrm{yba}}_{2}{\mathrm{cu}}_{3}{\mathrm{o}}_{y}$},}\ }\href {\doibase
  10.1103/PhysRevB.93.064513} {\bibfield  {journal} {\bibinfo  {journal} {Phys.
  Rev. B}\ }\textbf {\bibinfo {volume} {93}},\ \bibinfo {pages} {064513}
  (\bibinfo {year} {2016})}\BibitemShut {NoStop}%
\bibitem [{\citenamefont {Kim}\ \emph {et~al.}(2019)\citenamefont {Kim},
  \citenamefont {Coak}, \citenamefont {Baglo}, \citenamefont {Murphy},
  \citenamefont {Hill}, \citenamefont {Sutherland}, \citenamefont {Hatnean},
  \citenamefont {Balakrishnan},\ and\ \citenamefont {Park}}]{kim_2019}%
  \BibitemOpen
  \bibfield  {author} {\bibinfo {author} {\bibfnamefont {H.-L.}\ \bibnamefont
  {Kim}}, \bibinfo {author} {\bibfnamefont {M.~J.}\ \bibnamefont {Coak}},
  \bibinfo {author} {\bibfnamefont {J.~C.}\ \bibnamefont {Baglo}}, \bibinfo
  {author} {\bibfnamefont {K.}~\bibnamefont {Murphy}}, \bibinfo {author}
  {\bibfnamefont {R.~W.}\ \bibnamefont {Hill}}, \bibinfo {author}
  {\bibfnamefont {M.}~\bibnamefont {Sutherland}}, \bibinfo {author}
  {\bibfnamefont {M.~C.}\ \bibnamefont {Hatnean}}, \bibinfo {author}
  {\bibfnamefont {G.}~\bibnamefont {Balakrishnan}}, \ and\ \bibinfo {author}
  {\bibfnamefont {J.-G.}\ \bibnamefont {Park}},\ }\bibfield  {title} {\enquote
  {\bibinfo {title} {Modular thermal hall effect measurement setup for
  fast-turnaround screening of materials over wide temperature range using
  capacitive thermometry},}\ }\href {\doibase 10.1063/1.5108512} {\bibfield
  {journal} {\bibinfo  {journal} {Review of Scientific Instruments}\ }\textbf
  {\bibinfo {volume} {90}},\ \bibinfo {pages} {103904} (\bibinfo {year}
  {2019})}\BibitemShut {NoStop}%
\bibitem [{\citenamefont {Li}\ \emph {et~al.}(2017)\citenamefont {Li},
  \citenamefont {Xu}, \citenamefont {Ding}, \citenamefont {Wang}, \citenamefont
  {Shen}, \citenamefont {Lu}, \citenamefont {Zhu},\ and\ \citenamefont
  {Behnia}}]{Behnia_2017}%
  \BibitemOpen
  \bibfield  {author} {\bibinfo {author} {\bibfnamefont {X.}~\bibnamefont
  {Li}}, \bibinfo {author} {\bibfnamefont {L.}~\bibnamefont {Xu}}, \bibinfo
  {author} {\bibfnamefont {L.}~\bibnamefont {Ding}}, \bibinfo {author}
  {\bibfnamefont {J.}~\bibnamefont {Wang}}, \bibinfo {author} {\bibfnamefont
  {M.}~\bibnamefont {Shen}}, \bibinfo {author} {\bibfnamefont {X.}~\bibnamefont
  {Lu}}, \bibinfo {author} {\bibfnamefont {Z.}~\bibnamefont {Zhu}}, \ and\
  \bibinfo {author} {\bibfnamefont {K.}~\bibnamefont {Behnia}},\ }\bibfield
  {title} {\enquote {\bibinfo {title} {Anomalous nernst and righi-leduc effects
  in ${\mathrm{mn}}_{3}\mathrm{Sn}$: Berry curvature and entropy flow},}\
  }\href {\doibase 10.1103/PhysRevLett.119.056601} {\bibfield  {journal}
  {\bibinfo  {journal} {Phys. Rev. Lett.}\ }\textbf {\bibinfo {volume} {119}},\
  \bibinfo {pages} {056601} (\bibinfo {year} {2017})}\BibitemShut {NoStop}%
\bibitem [{\citenamefont {Xu}\ \emph {et~al.}(2020)\citenamefont {Xu},
  \citenamefont {Li}, \citenamefont {Lu}, \citenamefont {Collignon},
  \citenamefont {Fu}, \citenamefont {Koo}, \citenamefont {Fauqué},
  \citenamefont {Yan}, \citenamefont {Zhu}, \citenamefont {Behnia},\ and\
  \citenamefont {et~al.}}]{Behnia_2020}%
  \BibitemOpen
  \bibfield  {author} {\bibinfo {author} {\bibfnamefont {L.}~\bibnamefont
  {Xu}}, \bibinfo {author} {\bibfnamefont {X.}~\bibnamefont {Li}}, \bibinfo
  {author} {\bibfnamefont {X.}~\bibnamefont {Lu}}, \bibinfo {author}
  {\bibfnamefont {C.}~\bibnamefont {Collignon}}, \bibinfo {author}
  {\bibfnamefont {H.}~\bibnamefont {Fu}}, \bibinfo {author} {\bibfnamefont
  {J.}~\bibnamefont {Koo}}, \bibinfo {author} {\bibfnamefont {B.}~\bibnamefont
  {Fauqué}}, \bibinfo {author} {\bibfnamefont {B.}~\bibnamefont {Yan}},
  \bibinfo {author} {\bibfnamefont {Z.}~\bibnamefont {Zhu}}, \bibinfo {author}
  {\bibfnamefont {K.}~\bibnamefont {Behnia}}, \ and\ \bibinfo {author}
  {\bibnamefont {et~al.}},\ }\bibfield  {title} {\enquote {\bibinfo {title}
  {Finite-temperature violation of the anomalous transverse wiedemann-franz
  law},}\ }\href {\doibase 10.1126/sciadv.aaz3522} {\bibfield  {journal}
  {\bibinfo  {journal} {Science Advances}\ }\textbf {\bibinfo {volume} {6}}
  (\bibinfo {year} {2020}),\ 10.1126/sciadv.aaz3522}\BibitemShut {NoStop}%
\end{thebibliography}%

\appendix
\begin{section}{Longitudinal Boundary Term Approximation}\label{Dxy}

\begin{figure}
\centering
\includegraphics[width=.47\textwidth]{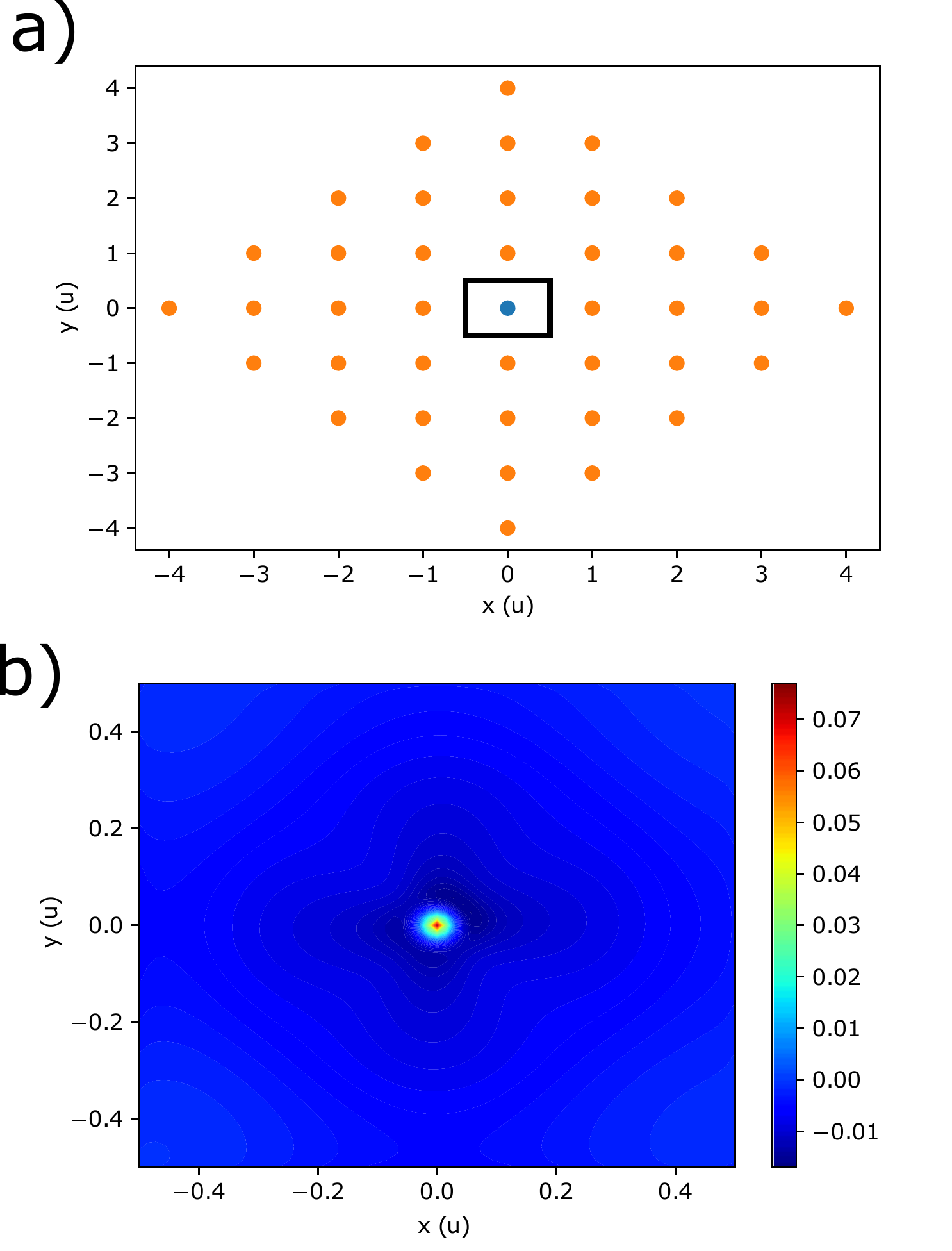}
\caption{ \label{BCSim} a) Lattice of mirror charges used to model boundary response for a zero-heat flow box. Note that there is an additional layer of mirror charges to handle the $T=0$ bottom boundary condition. b) The difference in simulated phase response (Fig. \ref{BSM}b) to the theoretical solution accounting for 3rd order mirror charges. Note the phase error appears to be due to finite beam size effects.}
\end{figure}

For the photothermal approach, the drive term $g$ was simulated as a narrow oscillatory Gaussian with width $\sigma$ centered at $(x_{0}, y_{0})$. The heating source exponentially decays with a skindepth $b$ from the top surface at $z_{0}$ and oscillates with frequency $\omega$, yielding
\begin{equation}
    g\left(\vec{x}, t\right) = Ge^{-\frac{\left(x - x_{0}\right)^2 + \left(y - y_{0}\right)^2}{2\sigma^2}}e^{\frac{z - z_{0}}{b}}e^{i\omega t}.
\end{equation}
In order to extract a two-dimensional phase observable, the real and imaginary parts of the simulated thermal response are projected into two dimensions with the same $e^{(z - z_{0})/b}$ weighting to approximate a probe laser and converted to a phase.

A closed-form expression for thermal boundary terms improves accuracy in $\textbf{D}$ for mesoscopic samples. If Eqn. \ref{baseTrue} is solved with $g = \delta(r'=0)$ as $T_{0}(r')$, a term of the form $T_{0}(r'')$ can be added to the boundary condition-free solution for any $r''$ centered outside of the domain of the sample. This allows for a mirror charge approach to solving boundary condition problems. For a $T=0$ boundary at $x = l$ and a source located at the origin, a ``mirror charge" centered at $x = 2l, y = z = 0$ can cancel out the boundary-condition free solution at $x = l$, giving an overall solution
\begin{equation}
     T = T\left(\frac{1}{r'}\exp\left[-\sqrt{i\omega}r'\right] - \frac{1}{r''}\exp\left[-\sqrt{i\omega}r''\right]\right).
\end{equation}
or
\begin{equation}
    T = T_{0}(r') - T_{0}(r'').
\end{equation}

Enforcing a condition of no heat flow across a boundary with norm $\hat{n}$, or $\textbf{D}\vec{\nabla}T \cdot \hat{n} = 0$, can also be solved with a mirror charge. Placing a source of the same sign as the original heating source at $x = 2l, y = z = 0$ cancels the heat flow across the $x = l$ boundary, giving an overall solution
\begin{equation}\label{gradBC}
    T = T\left(\frac{1}{r'}\exp\left[-\sqrt{i\omega}r'\right] + \frac{1}{r''}\exp\left[-\sqrt{i\omega}r''\right]\right)
\end{equation}
or
\begin{equation}
    T = T_{0}(r') + T_{0}(r'').
\end{equation}
This approach is seen in Fig. \ref{BCSim} and accounts for the observed phase up to finite beam size effects and simulation rounding error.

Transverse contributions can be analyzed beginning from Eqn. \ref{baseTrue}, adding a boundary at $x = l$, and no longer assuming $D_{xy} = 0$. The differential equation now reads
\begin{equation}
    i\omega T - \vec{\nabla}\cdot\left(\textbf{D}\Theta\left(l - x\right)\vec{\nabla}T\right) = g(r).
\end{equation}
Note the transverse diffusivity terms cancel inside the sample if $D_{xy} = -D_{yx}$. Evaluating the Heaviside derivative yields
\begin{equation}
    i\omega T - \vec{\nabla}\cdot\left(\textbf{D}\vec{\nabla}T\right) + \delta\left(x = l\right)\left[D_{xx}\frac{dT}{dx} + D_{xy}\frac{dT}{dy}\right] = g(r).
\end{equation}
The new boundary condition of no heat flow therefore is
\begin{equation}\label{BC}
    D_{xx}\frac{dT}{dx} = -D_{xy}\frac{dT}{dy}.
\end{equation}
For a discussion of the impact of irregular or rough boundaries, see Appendix \ref{rough}.

Eqn. \ref{BC} was solved for the $D_{xy} = 0$ in Eqn. \ref{gradBC}. Call the perturbation to that solution $T_{xy}$. Keeping only lowest order in $D_{xy}$, we find
\begin{equation}\label{LE}
    \frac{dT_{xy}}{dx} = -\frac{D_{xy}}{D_{xx}}\left(\frac{dT}{dy}\right) =  -2\frac{D_{xy}}{D_{xx}}\frac{dT_{0}}{dy}.
\end{equation}
In the far field approximation for two sources of the opposite sign separated by a distance $a \hat{x_i}$, the thermal response can be approximated as
\begin{equation}\label{LEp}
    T = T_{0}\left(r'_{1}\right) - T_{0}\left(r'_{2}\right) \approx \vec{\nabla}T_{0} \cdot a\hat{x_i} = \frac{dT_{0}}{dx_{i}}a.
\end{equation}
If the source response $T_{0}$ is replaced by a dipole response separated in the $x$-direction by $\delta x$, the resulting $T_{xy}$ is produced by a dipole separated in the $y$-direction by $\delta y$ situated at $x = 2l$. Plugging in dipoles, Eqn. \ref{LEp} is
\begin{equation}
    \delta y \frac{d^2T_{0}\left(r''\right)}{dxdy} = -2\frac{D_{xy}}{D_{xx}}\delta x \frac{d^2T_{0}\left(r'\right)}{dxdy}
\end{equation}
and is solved for 
\begin{equation}
    \delta y = 2\frac{D_{xy}}{D_{xx}}\delta x.
\end{equation}
Thus if the the heat source in the sample is an $x$-separated dipole, the boundary response is the mirror $x$-separated dipole and an additional $y$-separated dipole due to $D_{xy}$. The transverse diffusivity dipole is either decreased in source amplitude or separation relative to the original source dipole by $2 D_{xy}/D_{xx}$.

The dipole solution may be used to create an approximation of the $D_{xy}$ boundary response to a point source. Begin by superimposing a second dipole offset by the same $\delta x$ as the original source dipole. This yields a new source dipole separated by $2 \delta x$ and a boundary response of two identical $y$ separated dipoles at $x = 2l$ and $x = 2l + \delta x$. This process can be repeated an arbitrary number of times, leaving only one of the source dipole charges close to the boundary, and creating a boundary response of a series of identical $y$ separated dipoles stretching from $x = 2l$ to $x = \infty$. Approximate the boundary response $y$-separated dipoles as two lines of charges with density $\lambda_{h}$
\begin{equation}
    \lambda_h = \frac{2 A D_{xy}}{D_{xx}\delta x}
\end{equation}
for original source strength $A$. Approximating the dipole response by Eqn. \ref{LEp}, the temperature response is
\begin{equation}\label{initialEqn}
    T_{xy} = 2A\frac{D_{xy}}{D_{xx}}\frac{d}{dy}\int_{x_{0}}^{\infty} \frac{\exp[-\sqrt{-i\omega}u]}{u}dx_{l}
\end{equation}
where $u = \sqrt{(x - 2l)^2 + y^2 + z^2}$. However, Eqn. \ref{initialEqn} is not easily integrable to give a closed form solution of boundary response. Instead, such a closed form solution may be approximated by discretizing the lines of charge into two series of point sources. The length of the point source lines may also be reduced from infinity by considering the first order mirror reflection of lines of charge from opposing sides of a box, although in practice the boundary response was insensitive to changes in the line of charge length for lines of much longer length than the separation to the boundary.

\section{Crystal Shape and Roughness Simulation}\label{rough}

In order to handle oddly shaped samples such as crystals with rough boundaries or flakes, simulations on arbitrary mesh shapes must be developed. FEniCS allows for the construction of and simulation on meshes made of the superposition of fundamental shapes as seen in Fig. \ref{ArbShape}a. A two dimensional projection of this mesh can be created by choosing a random point in the 3D mesh and defining a minimum distance in the $x$-$y$ plane from that point. The $x$ and $y$ coordinates of the random point are stored in a 2D projection list, all coordinates within the minimum distance of that point are removed from the 3D mesh, and the maximum $z$-value in the list of close points is stored in a top surface list. This process is repeated until no points remain in the 3D mesh, defining a 2D projection and top surface for each projected point as seen in Fig. \ref{ArbShape}b. The phase can then be projected into a 2D observable using the known top surface and laser penetration depth. Using a triangulation of the 2D mesh points, the central location of the heating beam, and the assumption that phase monotonically decreases away from the beam center, a 2D phase unwrapping code can be defined on this triangulation by looking for $2\pi$ phase jumps stepping out along the triangulation from the heat source. The unwrapped phase with an asymmetric $\textbf{D}$ and strange sample shape is seen in Fig. \ref{ArbShape}c.

\begin{figure}
\centering
\includegraphics[width=\columnwidth]{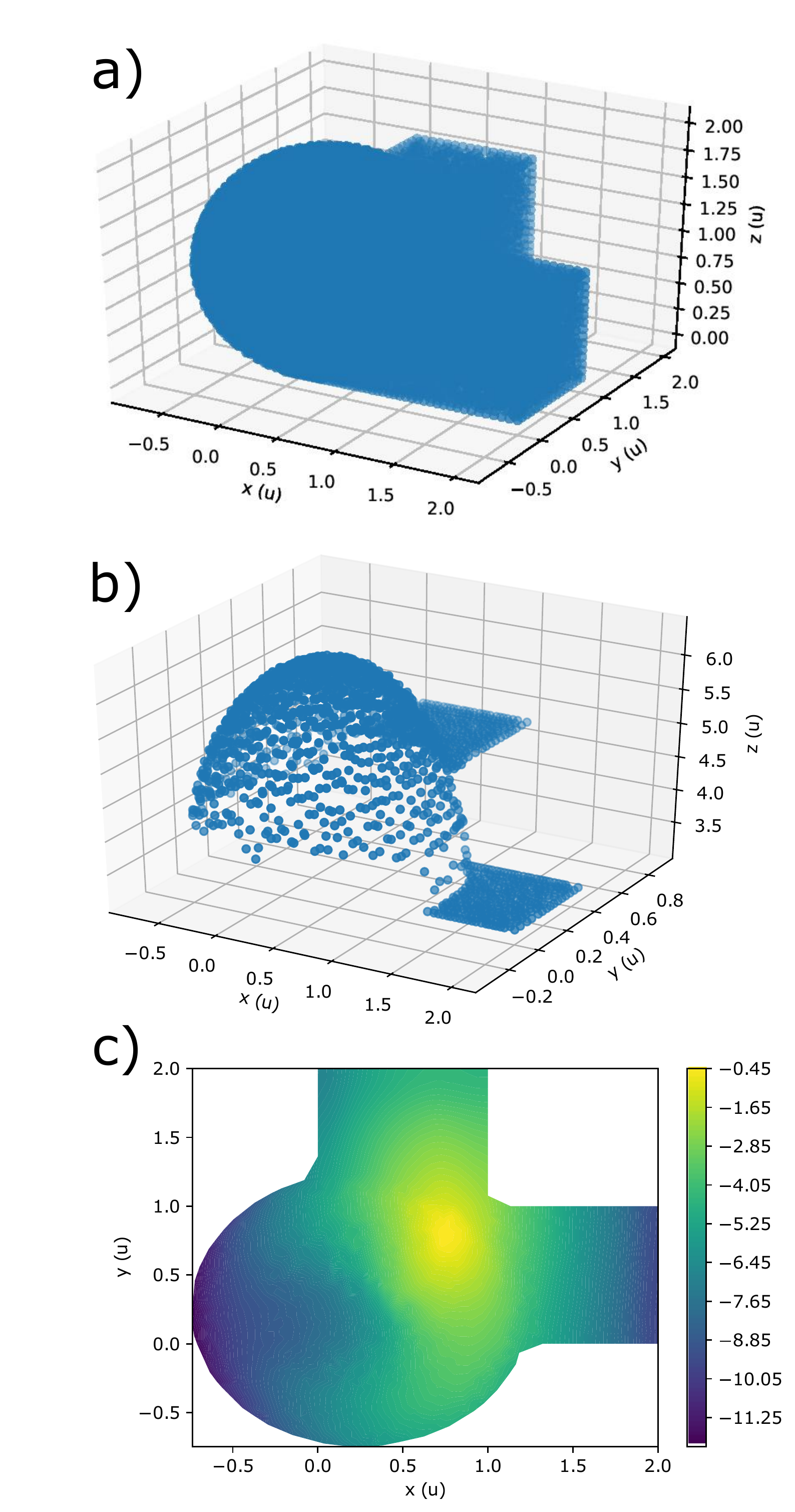}
\caption{ \label{ArbShape} a) Combined geometry of two boxes and a sphere, an arbitrary shape for simulation. b) The top surface of the geometry, used for phase projection into 2D. c) Projected 2D phase with $(D_{xx}, D_{yy}, D_{zz}) = (1, 5, 0.1)$~u$^2$/s.}
\end{figure}

As suggested by how $D_{xy}$ appears as a boundary term, it is easiest to measure $D_{xy}$ by the boundary. This raises concerns about if the form of the observed response may depend on microscopic details of the boundary. To investigate such an effect, the shift in phase due to $D_{xy}$ as a function of boundary roughness was investigated as seen in Fig. \ref{BCRough}. As seen in subfigures a-iv and b, the phase response from $D_{xy}$ is not limited to the boundary and peaks at $\sim \lambda/2$ away from the boundary. The general shape of the $D_{xy}$ response is also independent of the roughness scale $R$ as seen in subfigure b. The error due to surface roughness is also spacially limited to $R$ from the boundary until $R \sim \lambda$ as seen in Fig. \ref{BCRough}c. Overall, despite being a boundary effect, the $D_{xy}$ signal persists on the lengthscale $\lambda$ away from the boundary, and if $R << \lambda$, $D_{xy}$ should be a boundary-quality independent observable.

\begin{figure*}
\centering
\includegraphics[width=\textwidth]{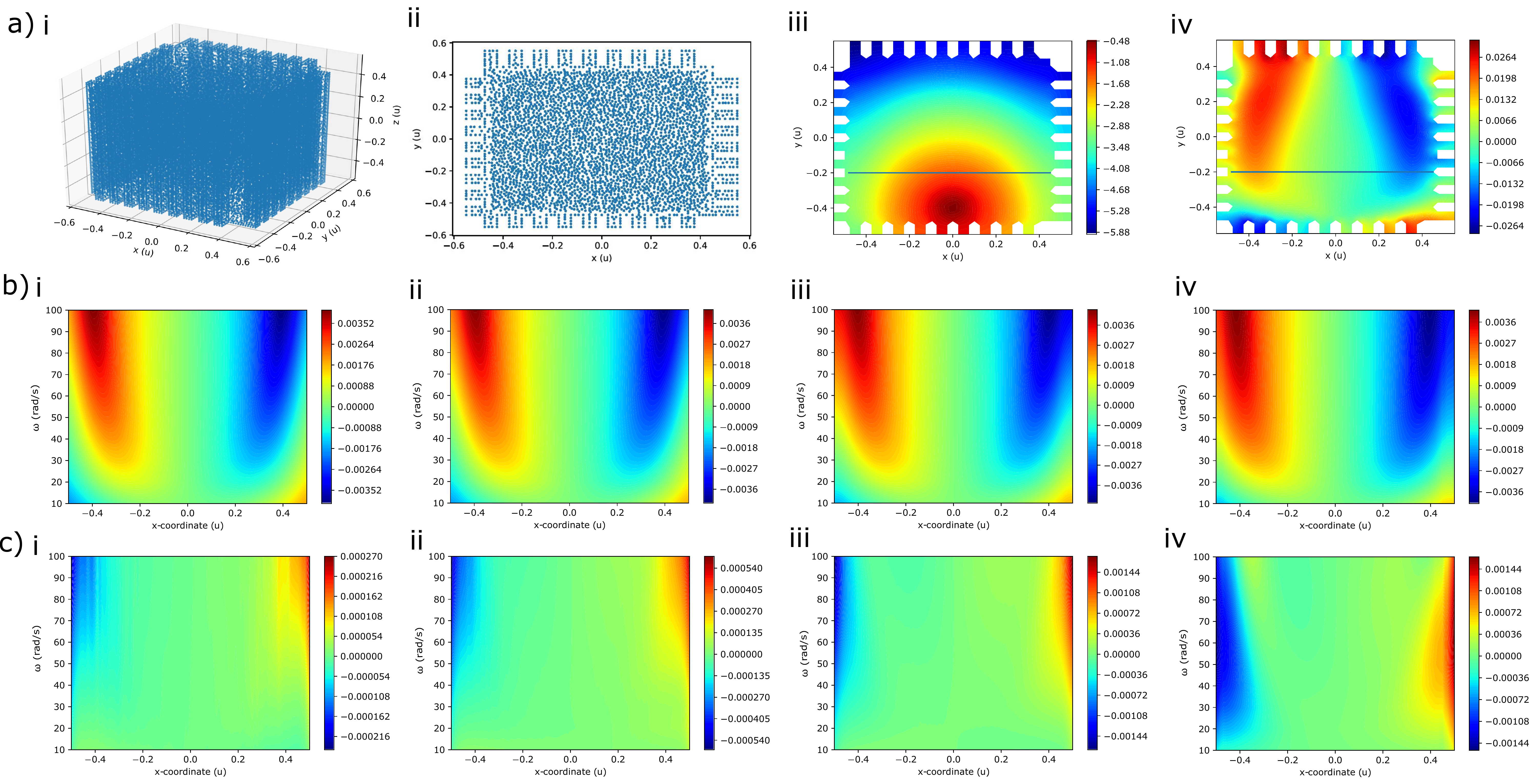}
\caption{ \label{BCRough} \textbf{a)} Background on the rough boundary condition simulation. The base box is 1x1x1~u and a roughness scale of 0.1~u is displayed. i) The full three dimensional roughened box. For a roughness scale of $R$, the teeth extend from the original box extent to $\pm R/2$ to maintain the same average box width and length. ii) A top-down two dimensional projection of the roughened box used to define the phase observed near the surface. iii) Resulting phase with $D_{xx} = D_{yy} = D_{zz} = 1$~u$^2$/s, $D_{xy} = .05$~u$^2$/s and $\omega = 50$~rad/s. iv) The difference in phase between $D_{xy} = 0$ and $D_{xy} = 0.05$~u$^2$/s. Note the black line at $y = -0.2$~u, the plots in parts b and c are generated from the difference in phase between $D_{xy} = 0$ and $D_{xy} = 0.01$ along this line as $\omega$ is changed. Note the mirror symmetry of the simulation result, used as a check of simulation fidelity. \newline \newline \textbf{b)} Difference in phase along the line at $y = -0.2$~u between $D_{xy} = 0$ and $D_{xy} = 0.01$~u$^2$/s as a function of $\omega$. Note the overall shape of the $D_{xy}$ response is only marginally changed as the roughness is increased.  \newline The roughness scale is changed between subfigures b i-iv. Roughness scale of i) 0.01~u, ii) 0.02~u, iii) 0.05~u, iv) 0.1~u. \newline \newline \textbf{c)} Difference in phase along the line at $y = -0.2$~u between $R = 0$ and nonzero roughness with $D_{xy} = 0.01$~u$^2$/s. The same roughness scales as in subfigure b are used. Note the changing scales in roughness-induced signal shift as roughness is increased. For $R = 0.01, 0.02, 0.05$~u, the difference in observed phase increases linearly with $R$. The roughness-induced shift saturates in amplitude about $R = 0.05$~u, but begins to appear in lower $\omega$. In the linear response regime, the amplitude of roughness-induced shift is much smaller than the observed $\delta \phi$ due to $D_{xy}$ and is bound to the boundary, while the peak $D_{xy}$ signal is $\sim \lambda/2$ away from the boundary.}
\end{figure*}

\end{section}

\end{document}